%% file: 1-birefringence_2022-0508-ver1.tex
\UseRawInputEncoding

\documentclass[
preprint, 
preprintnumbers,superscriptaddress,showpacs,supberscriptaddress,
nofootinbib,longbibliography]{revtex4-1}

\usepackage[
bookmarks=true,colorlinks,linkcolor=blue,urlcolor=cyan,citecolor=red]{hyperref}

\input{preamble}
\graphicspath{{./figs/}}

\newcommand{\ra}{{ra}}
\newcommand{\rr}{{rr}}
\newcommand{\ar}{{ar}}

\newcommand{\temp}{{\rm med}}
\newcommand{\Db}{{\rm D}}

\newcommand{\arctanh}{{\rm arctanh}}

\begin{document}


\title{In-medium polarization tensor in strong magnetic fields (I): 
\\
Magneto-birefringence at finite temperature and density}

\author{Koichi Hattori}
\affiliation{Zhejiang Institute of Modern Physics, Department of Physics, 
Zhejiang University, Hangzhou, 310027, China}
\affiliation{Research Center for Nuclear Physics, Osaka University, Osaka 567-0047 Japan.}

\author{Kazunori Itakura}
\affiliation{Nagasaki Institute of Applied Science (NiAS), 536 Aba-machi, 
Nagasaki-shi, Nagasaki, 851-0193, JAPAN}
\affiliation{High-Energy Accelerator Research Organization (KEK), 1-1 Oho, 
Tsukuba, Ibaraki, 305-0801, JAPAN}


\begin{abstract}

We investigate in-medium polarization effects of the fermion and antifermion pairs 
at finite temperature and density in strong magnetic fields within the lowest Landau level approximation. 
Inspecting the integral representation of the polarization tensor by analytic and numerical methods, 
we provide both the real and imaginary parts of the polarization tensor 
obtained after delicate interplay between the vacuum and medium contributions 
essentially due to the Pauli-blocking effect. 
Especially, we provide a complete analytic form of the polarization tensor 
at zero temperature and finite density that exhibits an exact cancellation and associated relocation 
of the singular threshold behaviors for a single photon decay to a fermion and antifermion pair. 
As a physical application of the in-medium polarization tensor, we discuss the magneto-birefringence 
that is polarization-dependent dispersion relations of photons induced by the strong magnetic fields. 



\end{abstract}

\maketitle
\tableofcontents

\section{Introduction}

Vacuum states under strong electromagnetic fields have been intensively studied 
since Heisenberg and Euler established an effective action 
with effects of the strong fields \cite{Heisenberg:1935qt}. 
One of the prominent consequences of the extremely strong fields 
may be the particle production in strong electric fields \cite{sauter1931behavior, 
Heisenberg:1935qt}, which is often called the Schwinger mechanism \cite{Schwinger:1951nm}. 
The particle production is induced by the energy transfer from the electric field 
to quantum fluctuations of electrically charged particles in vacuum. 
The Schwinger mechanism is inherently a dynamical process, and has been discussed 
as the mechanisms for formation of the quark-gluon plasma 
(see, e.g., Ref.~\cite{Gelis:2015kya, Taya:2017pdp, Berges:2020fwq} for reviews). 
The framework has been extended to particle creation in time-dependent gravitational fields 
that can be important for explaining striking aspects of the primordial universe 
\cite{Parker:1971pt, Martin:2007bw}. 
On the other hand, the Lorentz force does not do work on charged particles, 
so that there is no energy transfer from (constant) magnetic fields to charged particles. 
One may ask whether the static vacuum state under the strong magnetic fields 
leave any physical consequences.

Magnetic fields induce quantization in the energy spectrum of charged particles 
i.e., the Landau quantization \cite{Landau1930}. 
The Landau quantization has been known to play an important role 
in various phenomena such as magnetization and transport phenomena 
in condensed matter physics that can be probed as responses to external perturbations. 
The Landau quantization should occur to the quantum fluctuations in vacuum as well, 
and give rise to physical consequences when the quantized vacuum fluctuations 
respond to external probes that can be, for example, perturbative electric fields 
and propagating photons like the cases of quantum Hall systems and optics, respectively.

The coupling between the external electromagnetic probes and 
the charged fluctuations in magnetic fields can be captured by a vector-current correlator. 
Within QED, the one-loop vacuum polarization diagram should work as 
a good approximation to the correlator. 
However, the charged particles on the loop can interact with the magnetic fields 
multiple times when the magnetic field is strong enough. 
Intuitively speaking, this occurs when the Lorentz force is strong enough to 
form a closed cyclotron orbit within the lifetime of virtual fluctuations. 
Then, the one-loop polarization tensor needs to be resummed with respect to 
insertion of the external lines for the magnetic field 
and thus acquires a nonlinear dependence on the magnetic-field strength. 
This is the quantum and nonlinear regime where the Landau quantization 
manifests itself in physical quantities.

An important consequence of the vacuum fluctuations in the strong magnetic fields is 
a polarization-dependent refractive index of photons propagating 
in the ``magnetized vacuum,'' which is called the vacuum birefringence \cite{Toll:1952rq}. 
There are a number of experimental efforts to probe the vacuum birefringence 
especially in laser physics \cite{DiPiazza:2011tq, DellaValle:2015xxa, Fedotov:2022ely} 
and neutron-star physics \cite{Harding:2006qn, Mignani:2016fwz, Enoto:2019vcg}, 
which has been one of the main subjects in strong-field QED. 
Another consequence is the axial-charge generation in response to a (perturbative) electric field 
that includes a contribution from the chiral anomaly \cite{Adler:1969gk, Bell:1969ts}. 
The aforementioned two-point vector-vector correlator can capture the axial-charge generation 
because the electric-charge separation by the electric field implies the axial-charge separation 
due to the chirality-momentum locking in the presence of the Landau quantization \cite{Nielsen:1983rb}. 
Technically, the vector-vector and vector-axial vector correlators are connected with each other 
via an identity between the vector and axial-vector currents. 
Various aspects of those two interesting vacuum phenomena have been studied.

However, in real systems, it is often important to include medium effects at finite temperature and/or density. 
In this series of papers, we thus investigate (i) the photon properties 
and (ii) the axial-charge generation induced by polarization of the medium particles 
as well as the vacuum fluctuations under strong magnetic fields. 
We inspect the medium contribution to the one-loop polarization diagram with thermal field theory 
by both analytic and numerical methods. 
The integral representation of the polarization tensor was derived 
within the lowest Landau level (LLL) approximation \cite{Fukushima:2015wck}. 
More recently, the polarization tensor was investigated in all the Landau levels 
focusing on the imaginary part  \cite{Wang:2020dsr, Wang:2021ebh, Wang:2021eud}. 
We find that, in case of the fermion fluctuations, 
the sum of the vacuum and medium contributions is not simply constructive 
but is rather destructive in some kinetics basically due to the Pauli-blocking effect 
on the thermally occupied positive-energy states. 
It is also important to note that the vacuum contribution is comparable 
in magnitude to the thermal contribution even at high-temperature and/or -density limit 
unlike in the hard thermal loop approximation in (3+1) dimensions, 
giving rise to delicate interplay between the vacuum and thermal contributions. 
This is due to an effective dimensional reduction to (1+1) dimensions in strong magnetic fields. 
It is worth mentioning the related works 
in (1+1) dimensional QED \cite{Dolan:1973qd, Baier:1991gg, Smilga:1991xa, Kao:1998yt}.

In this first paper of the series, we first summarize computation of the medium-induced polarization tensor 
within the LLL approximation 
so that one can grab a qualitative picture of medium effects. 
This may be best achieved by investigating the imaginary part of the polarization tensor 
that provides the squared amplitude of the on-shell processes. 
We provide an analytic expression of the imaginary part 
and discuss the on-shell kinematics of the relevant physical processes. 
We find that the medium effects suppress the decay rate of a single photon to 
a fermion and antifermion pair\footnote{
Note that this 1-to-2 on-shell process is kinematically allowed in magnetic fields 
for real photons as well as virtual photons as a consequence of the Landau quantization 
(see, e.g., Refs.~\cite{Hattori:2012je, Hattori:2012ny} and references therein). 
} due to the Pauli-blocking effect 
and also open a new reaction channel known as the Landau damping which is absent in vacuum.  
We then investigate the real part of the polarization tensor both by analytic and numerical methods. 
The real and imaginary parts are entangled with each other through the dispersion integral. 
We provide a complete analytic expression at finite density and zero temperature 
and show that the threshold energy of the single-photon decay is shifted from 
the fermion pair mass to the chemical potential 
because the vacant final states are only available above the Fermi surface. 
The analytic expression clearly verifies occurrence of the threshold shift.

Finally, we investigate the photon dispersion relation obtained 
from the pole position of the photon propagator 
with the polarization tensor resummed as a geometrical series. 
Only the photon polarization mode oscillating in parallel to the magnetic field 
can couple to the charged particles in the LLL due to shrinking of the cyclotron orbit, 
and acquires a nontrivial dispersion relation. 
This polarization dependence is nothing but the magnetic-field-induced birefringence. 
Specifically, we compute the photon masses by taking the vanishing photon frequency 
and momentum limits. 
It should be noticed that there are two different photon masses 
since the limits do not commute with each other in general (cf. Ref.~\cite{Bellac:2011kqa}). 
Taking the vanishing momentum limit first gives an energy gap of the photon dispersion relation 
as known as the plasma frequency 
that characterizes collective excitations in the medium. 
The other order of the limits, where the vanishing frequency is taken first, 
gives the Debye screening mass that characterizes the screening length of a static charge. 
The screening effect has been also investigated 
by the recent lattice QCD simulations \cite{Bonati:2014ksa, Bonati:2016kxj, Bonati:2017uvz}.

What is interesting in the LLL is that photons acquire a finite mass even without the medium effects 
in analogy with the Schwinger mass \cite{Schwinger:1962tn, Schwinger:1962tp}, 
because the dynamics in the strong-field limit, where the (classical) cyclotron radius vanishes, 
is (1+1) dimensional one along the magnetic-field direction. 
We investigate interplay between the vacuum and medium contributions. 
We provide consistency checks at infinite temperature or density limit 
and confirm that the Schwinger mass is reproduced in those limits. 
Finally, it should be noticed that the LLL approximation generally works in the strong magnetic fields, 
where the excitations to the higher Landau levels are highly suppressed, 
requiring that the magnetic field strength (multiplied by the coupling constant) be much larger than 
the temperature and chemical potential, i.e., $ |e B| \gg T, \, \mu $. 
Thus, the above limits should be understood as a formal limit.

This paper is organized as follows. 
In Sec.~\ref{sec:diagrams}, we provide the vacuum and medium contributions 
to the one-loop polarization tensor 
and an explicit form of the photon propagator with the polarization tensor resummed. 
The medium contribution is given in a one-dimensional integral form. 
In Sec.~\ref{sec:medium}, we inspect the medium contribution by performing the remaining integral. 
In Sec.~\ref{sec:screening}, we discuss the photon masses as a physical application. 
Another application to the axial-charge generation is left to 
the second paper of the series. 
We then give discussions and summary of this paper, 
followed by several appendices for line-by-line explanations of the analytic calculations.

Throughout the paper, we focus on a single-flavor Dirac fermion for simplicity, 
and introduce a single chemical potential conjugate to its number density. 
We also assume that a constant magnetic field is applied along the third spatial direction 
without loosing generality. 
Accordingly, we use metric conventions $ g^{\mu\nu} = \diag ( 1, -1, -1, -1) $, 
$ g_\para^{\mu\nu} = \diag(1, 0,0,-1) $, and $ g_\perp^{\mu\nu} = (0,-1,-1,0) $ 
and associated notations $ q_\para^2 := q_\mu q_\nu g_\para^{\mu\nu} $ 
and $ q_\perp^2 := q_\mu q_\nu g_\perp^{\mu\nu} = - |\bq_\perp|^2$ for four vectors $ q^\mu $.




\section{Polarization tensors}

\label{sec:diagrams}

In this section, we summarize computation of 
the one-loop polarization tensors in a self-contained way. 
Since magnetic fields do not directly couple to photons, 
they affect the one-loop polarization tensors indirectly through 
the coupling to the electric-charge flow on the loop. 
Constant magnetic fields induce cyclotron motions of charged particles at the classical level, 
and such cyclic motions result in the Landau quantization 
in a similar manner to the quantization of harmonic oscillators. 
In addition, magnetic fields give rise to the Zeeman effect. 
We focus on the strong magnetic field limit where contributions to the polarization tensors 
are dominated by fermions in the ground state that is the lowest Landau level (LLL) 
with spin polarized along the magnetic field.

\subsection{Fermion propagator in strong magnetic fields}

To compute the polarization tensors, we first provide the fermion propagator 
in strong magnetic fields. 
There are various equivalent ways to get the propagator 
(see, e.g., Refs.~\cite{Schwinger:1951nm, Brown:1975bc, Dittrich:1975au, 
Dittrich:1985yb, Chodos:1990vv, Gusynin:1995nb, Hayata:2013sea,Miransky:2015ava, 
Dittrich:2000zu, Hattori:2020guh}).  
The Landau quantization manifests itself in the fermion propagator 
as a result of coherent interaction between the magnetic field and fermions. 
We provide a brief derivation focusing on the LLL 
on the basis of the Ritus-basis formalism~\cite{Ritus:1972ky, Ritus:1978cj} 
(see also a review part in Ref.~\cite{Hattori:2020htm}).

We start with the Dirac equation 
\begin{align}
	\left( i \slashed D - m \right) \psi = 0 \, , \label{eq:Dirac}
\end{align}
where the covariant derivative in an external magnetic field is given as 
\begin{align}
	D^\mu  \equiv \partial ^\mu + i q_f A^\mu \, . \label{eq:covariantD-QED}
\end{align}
The electric charge $q_f$ takes a positive (negative) value for positively (negatively) charged fermions. 
The gauge field $ A^\mu $ is for a constant magnetic field. 
Without loosing generality, we assume that the constant magnetic field 
is oriented in the third spatial direction. 
Then, we have a commutator
\begin{align}
	[ i D^1 , iD^2] =  iq_f B\, ,
\end{align}
with the other components vanishing. 
Motivated by this commutator, we introduce ``creation and annihilation operators'' as 
\begin{align}\label{eq:aadagger}
	\hat a \equiv \frac{1}{ \sqrt{2 | q_fB|} } ( i D^1 - s_f  D^2 )\ ,\quad 
	\hat a^\dagger \equiv \frac{1}{ \sqrt{2 |q_f B|} } ( i D^1 + s_f D^2 )
	\, ,
\end{align}
where $ s_f = \sgn(q_f B) $. These operators satisfy the commutation relations 
$  [\hat{a}, \hat{a}^\dagger] =1 $ and $[\hat{a}, \hat{a}] = [\hat{a}^\dagger, \hat{a}^\dagger] = 0$. 
The state of the LLL is ``annihilated'' as $ \hat a | 0\rangle =0 $.

We would like to get Green's function $ S(x^\mu ,x ^{\prime\mu} ) $ of the Dirac operator such that 
\begin{eqnarray}
\label{eq:Dirac-S}
(i \slashed D _x -m )  S(x^\mu ,x ^{\prime\mu}) = i \delta^{(4)}(x^\mu -x ^{\prime\mu})
 \, .
\end{eqnarray}
Notice that the Dirac operator can be rewritten as 
\begin{eqnarray}
i \slashed D  -m 
\= ( i \slashed \partial _\parallel -m) 
- \sqrt{2|q_f B|}  \gamma^1\left(\hat a \prj_+ + \hat a^\dagger \prj_- \right) 
 \, ,
 \label{Dirac_Ritus}
\end{eqnarray}
with the spin-projection operator $ \prj _\pm 
=  \left(1 \pm  i s_f  \gamma^1 \gamma^2  \right)/2 $. 
We now project the propagator onto the LLL wave function 
\begin{eqnarray}
S(x^\mu ,x ^{\prime\mu}) \= \int \frac{d^2p_\para}{(2\pi)^2} \int \frac{dp_y}{2\pi} 
e^{-ip_\para \cdot (x_\para-x'_\para) + i p_y (y-y')}
\nnb
&&
\times
\phi_\LLL (x - \frac{ p_y}{q_fB} )  {\mathcal S}_\LLL(p_\para) 
\phi^\ast_\LLL  (x - \frac{ p_y}{q_fB} )  \prj_+
\, .
\end{eqnarray} 
We denote the first and second components of the four vector $  x^\mu$ 
as $ x $ and $y  $, respectively. 
We have chosen the Landau gauge $ A^\mu = (0, 0, Bx ,0) $ for the external magnetic field. 
Then, the second component of the fermion momentum $ p_y := p^2$ is still a good quantum number, 
and the plane wave $ e^{ i p_y y} $ serves as the eigenfunction of the Dirac operator. 
This also suggests that the LLL states can be labelled by $ p_y  $ as $   | 0, p_y \rangle  $ 
with the lowest principal quantum number $  \hat a^\dagger \hat a | 0, p_y \rangle = 0  $ 
and the normalization $ \langle 0, p_y  | 0, p'_y \rangle  = \delta(p_y - p_y') $. 
The rest part of the LLL wave function is given by the Hermite polynomial 
(see, e.g., Appendix in Ref.~\cite{Hattori:2020htm}) 
\begin{eqnarray}
e^{ i p_y y}  \phi_\LLL  (x - \frac{ p_y}{q_fB} )  := \langle x^\mu_\perp | 0, p_y \rangle 
= e^{ i p_y y}  \Big( \frac{ |q_f B| }{ \pi } \Big)^{\frac14} e^{ -  \frac{ |q_fB| }{2} (x - \frac{ p_y}{q_fB} )^2  }
\, .
\end{eqnarray} 
We will shortly clarify how the gauge dependence of the wave function is made tractable. 
The propagator takes a simple form in this basis called 
the Ritus basis \cite{Ritus:1972ky, Ritus:1978cj}, 
just because it is spanned by the eigenfunctions of the Dirac operator. 
After the LLL projection, we find 
\begin{eqnarray}
{\mathcal S}_\LLL(p_\para)  = \frac{i}{ \sla p_\para - m}
\, .
\end{eqnarray}
This serves as Green's function of Eq.~(\ref{eq:Dirac-S}), 
assuming that the completeness is here saturated by the (degenerate) ground states 
\begin{eqnarray}
\delta^{(2)} (x_\perp^\mu - x^{\prime\mu}_\perp) = \sum_{n=0}^\infty \int \frac{dp_y}{2\pi} 
 \langle x^\mu_\perp | n, p_y \rangle \langle n, p_y | x^{\prime\mu}_\perp \rangle
\sim   \int \frac{dp_y}{2\pi}  \langle x^\mu_\perp | 0, p_y \rangle \langle 0, p_y | x^{\prime\mu}_\perp \rangle
\, .
\end{eqnarray}
If one includes all the higher Landau levels (hLLs), 
they provide additive contributions to the fermion propagator 
that is given by summation over the Landau levels (see, e.g., Ref.~\cite{Gusynin:1995nb}).

The Ritus basis, however, is not an eigenfunction of dynamical photon fields. 
Thus, it is equally useful to get the LLL propagator in the Fourier basis 
so that the perturbative interaction vertex with a photon field is simply given by 
the delta functions for the four-momentum conservation as in usual perturbation theories. 
After some arrangements, we get 
\begin{eqnarray}
S(x^\mu ,x ^{\prime\mu}) \= 
e^{  i \Phi_A(x_\perp^\mu ,x_\perp ^{\prime\mu})  }   \int \frac{d^2p_\para}{(2\pi)^2} \int \frac{dp'_y}{2\pi} 
e^{-ip_\para \cdot (x_\para-x'_\para) + i p'_y (y-y')}
\nnb
&& \times
\phi_\LLL  ( \frac{ x - x'}{2} - \frac{ p'_y}{q_fB} ) {\mathcal S}_\LLL(p_\para) 
\phi^\ast_\LLL  ( -  \frac{ x - x'}{2}  - \frac{ p'_y}{q_fB} )  \prj_+
\nnb
\= e^{  i \Phi_A(x_\perp^\mu ,x _\perp^{\prime\mu})  }  \int \frac{d^4p}{(2\pi)^4} e^{ - i p \cdot x}
\Big[\, 2 e^{  -  \frac{ |\bp|^2 }{|q_fB|}    }   {\mathcal S}_\LLL(p_\para)  \prj_+ \, \Big]
\, ,
\end{eqnarray}
where $ \Phi_A(x_\perp^\mu ,x_\perp ^{\prime\mu}) :=  \frac{ q_f B}{2}  (x + x)  (y-y') $. 
In the first line, the integral variable was shifted as $ p_y \to p_y' = p_y -  (x + x') q_f B/2  $, 
making the integrand translation invariant. 
The Fourier representation of the translation-invariant part is given as 
\begin{eqnarray}
S_\LLL(p) := 2 e^{  -  \frac{ |\bp|^2 }{|q_fB|}    }   {\mathcal S}_\LLL(p_\para)  \prj_+
= 2  e^{ -\frac{|\bp_\perp|^2}{|q_fB|} } \frac{ i }{ \sla p_\parallel - m  } \prj_+
\label{eq:prop-LLL}
\, .
\end{eqnarray}  
The phase $ \Phi_A $ is a gauge-dependent quantity 
called the Schwinger phase \cite{Schwinger:1951nm}. 
In case of the Landau gauge employed here, the gauge configuration breaks 
the translational invariance in the $  x$ direction as captured by this phase. 
The rest part of $ S(x^\mu ,x ^{\prime\mu})  $ is gauge invariant as well as translation invariant, 
and we were able to transform it to the Fourier basis with a single momentum $ p^\mu $. 
One can keep track of the gauge dependence with the Schwinger phase $ \Phi_A $ 
that should go away in finial results of gauge-invariant quantities. 
When we compute the fermion one-loop in the following sections, 
the Schwinger phase vanishes as $ \Phi_A(x_\perp^\mu ,x_\perp ^{\prime\mu}) 
+ \Phi_A(x_\perp^{\prime\mu} ,x _\perp^{\mu}) =0 $, 
indicating a gauge invariance of the one-loop polarization diagrams. 
Therefore, we do not write the Schwinger phases explicitly hereafter.

%
%
%
%
%

\subsection{Vacuum contribution}

\label{sec:VP_vac}

By using the LLL propagator (\ref{eq:prop-LLL}), 
one can write down the one-loop vacuum polarization tensor 
\begin{eqnarray}
i \Pi_\vac^{\mu\nu} (q) =  - ( - i q_f)^2 \int \!\! \frac{d^4p}{(2\pi)^4} 
\tr [ \gam^\mu S_\LLL(p) \gam^\nu S_\LLL(p+q) ]
\, .
\end{eqnarray}
One can perform the Dirac trace and the integral straightforwardly, 
taking care of the superficial ultraviolet divergence. 
After an appropriate regularization by, e.g., the dimensional regularization 
described in Appendix~\ref{sec:comp-vacuum}, 
we get the vacuum polarization tensor 
\begin{eqnarray}
&& \Pi_\vac^{\mu\nu} = m_B^2 
e^{-\frac{\vert \bq_\perp \vert^2}{2 \vert q_fB\vert}} \tilde  \Pi_\para ^\vac \prj_\para^{\mu\nu} 
\label{eq:transverse-op}
\, ,
\end{eqnarray}
where the gauge-invariant tensor structure and 
the vacuum contribution to the scalar function $\tilde  \Pi_\para^\vac $ are, respectively, given as 
\begin{subequations}
\begin{eqnarray}
&& \prj_\para^{\mu\nu}  =  g^{\mu\nu}_\para - \frac{ q_\para^\mu q_\para^\nu}{q_\para^2} 
\, ,
\\
&&\tilde  \Pi_\para^\vac  (q_\para^2) =  1- I \big( \frac{q_\parallel^2}{4m^2} \big) 
\label{eq:vac_massive}
\, ,
\end{eqnarray}
\end{subequations}
with $m_B^2 = |q_fB|/(2\pi) \cdot q_f^2/\pi  $. 
Since $ m_B^2  $ is a dimension-two quantity, 
the $ I $ function can only be a function of a dimensionless variable: 
\begin{eqnarray}
I( x + i \omega \epsilon) = 
\left\{
\begin{array}{ll}
\frac{1}{2} \frac{1}{ \sqrt{x(x-1) } }
\ln  \frac{ \sqrt{x(x-1) } - x }{  \sqrt{x(x-1) } + x } 
& x < 0 
\\
\frac{1}{ \sqrt{x(1 - x) } }\arctan \frac{x } { \sqrt{x(1-x) } } 
& 0 \leq x < 1
\\
\frac{1}{2} \frac{1}{ \sqrt{x(x-1) } }
\left[ \,\ln  \frac{x -  \sqrt{x(x-1) } }{x +  \sqrt{x(x-1) } } 
+ i \, \sgn(\omega)  \pi \, \right]
& 1 \leq x
\end{array}
\right.
\label{eq:I}
\, ,
\end{eqnarray} 
were we made use of the analytic continuation according to 
a relation, $ \arctan z = \frac{i}{2} \ln  \frac{ 1 - i z }{ 1 + i z} $. 
We took the retarded prescription $ \omega \to \omega + i\epsilon $ that 
yields an infinitesimal displacement $ q_\para^2 \to q_\para^2 + i \omega \epsilon $.\footnote{
We used a relation $ \arctan \frac{x } { \sqrt{x(1-x) } }  
= \frac{i}{2} \ln \frac{\sqrt{x(x-1) }  -  x}{\sqrt{x(x-1) }  +  x} $ when $ x < 0 $ or $ 1 \leq x  $. 
When $ x\geq 1 $, the argument of the logarithm takes a negative value, 
so that the logarithm becomes a complex-valued function, $ \arctan \frac{x } { \sqrt{x(1-x) } }  
= \frac{i}{2} [ \ln \frac{ |\sqrt{x(x-1) }  -  x|}{\sqrt{x(x-1) }  +  x} + i  \sgn( \omega) \pi ] $. 
The sign of the imaginary part depends on that of the infinitesimal displacement. 
When $ x < 0 $, the argument of the logarithm is positive definite, i.e., 
$ \frac{ \sqrt{x(x-1) } - x }{  \sqrt{x(x-1) } + x }  = \frac{ - x }{ ( \sqrt{x(x-1) } + x)^2 }  \geq0  $. 
}
The advanced and causal correlators can be computed in the same manner. 
One can find two limiting behaviors 
\begin{eqnarray}
\label{eq:I-limits}
I(0) = 1 \, , \quad I(\infty) = 0
\, .
\end{eqnarray}
Therefore, in the infinite mass limit $  q_\para^2/m^2 \to 0$, 
we simply find that $\tilde  \Pi_\para^\vac  \to 0 $ 
because the fermion excitations are suppressed. 
In the massless limit $  q_\para^2/m^2 \to \infty$, we have
\begin{eqnarray}
\tilde  \Pi_\para^\vac  (q_\para^2) \=  1
\label{eq:Pi-massless}
\, .
\end{eqnarray}

Here are a couple of comments in order. 
The factor of $ |q_fB|/(2\pi)  $ in $m_B^2  $ is called the Landau degeneracy factor 
that accounts for the density of degenerate states in an energy level in the Landau quantization. 
The degeneracy originates from the translational symmetry in the plane transverse to 
the magnetic-field direction; there is no energetically preferred position 
for the center coordinate of cyclotron motion. 
The rest of the factor $ q_f^2/\pi  $ is known as the Schwinger mass 
in the (1+1) dimensional massless QED \cite{Schwinger:1962tn, Schwinger:1962tp} 
that provides a photon mass in the gauge-invariant way. 
This is made possible by the presence of the $ q_\para^2 $ pole 
in Eq.~(\ref{eq:transverse-op}) in the massless case. 
Here, the Schwinger mass appears in the (3+1)-dimensional QED 
because of the factorization of the longitudinal and transverse 
subspaces due to the Landau quantization, leading to the dimensional reduction. 
In this paper, we refer to $m_B^2  $ as the Schwinger mass for simplicity 
including the Landau degeneracy factor. 
We will inspect photon masses induced by medium contributions later in this paper.

Lastly, the imaginary part in Eq.~(\ref{eq:I}) indicates occurrence of 
a fermion-antifermion pair creation when the photon momentum satisfies 
the threshold condition $ q_\para^2 \geq (2m)^2 $ that is the invariant mass of a pair in the LLL states. 
In the massless limit (\ref{eq:Pi-massless}), 
the scalar function $ \tilde  \Pi_\para^\vac  (q_\para^2) $ does not yield any imaginary part. 
An imaginary part only comes from the $ q_\para^2 $ pole factorized 
as the tensor part $ \prj_\para^{\mu\nu} $ 
in the present notations, and is proportional to the delta function $ \delta(q_\para^2) $, 
meaning that there is no (invariant) momentum transfer from a photon to a fermion-antifermion pair. 
Such a pair creation does not usually happen since a hole state in the Dirac sea needs to be 
created by a {\it diabatic} transition of a particle from the negative- to positive-energy branches. 
The pair creation in an {\it adiabatic} process can occur only because 
the positive- and negative-energy states are directly connected with each other 
in the (1+1)-dimensional gapless dispersion relation.\footnote{
On the other hand, the diabatic pair creation is prohibited in the (1+1)-dimensional massless case 
due to the absence of chirality mixing. 
If a fermion and antifermion pair were created, they should have the same helicity 
in the center-of-momentum frame of the pair, meaning that they belong to 
different chirality eigenstates in strict massless theories. 
A similar prohibition mechanism is known as the ``helicity suppression'' 
in the leptonic decay of charged pions \cite{Donoghue:1992dd, Zyla:2020zbs}. 
} 
Namely, the pair creation is interpreted as the spectral flow along the linear dispersion relation 
that occurs independently in the right- and left-handed sectors, leading to the chiral anomaly. 
Finite mass effects, which are captured by the $ I $ function, modify the mechanism of the pair creation. 
In the second paper of the series, 
we will discuss the axial Ward identity that contains not only the anomalous term 
but also the pseudoscalar condensate term. 
The latter depends on the fermion mass as well as temperature/density, 
so that the axial Ward identity also depend on those parameters as a whole.


\subsection{Medium contribution at finite temperature and density}

\label{sec:VP_therm}

We summarize the medium contribution to the polarization tensor 
at finite temperature and density from the real-time formalism. 
The LLL propagator is given in Eq.~(\ref{eq:prop-LLL}) in vacuum. 
In the r-a basis of the real-time formalism, this propagator is extended to the following matrix form 
\begin{subequations}
\label{eq:S-all}
\begin{eqnarray}
\label{eq:Sra}
 S_{ra}(p) &=& 2 i\, e^{- \frac{ \vert \bm p_\perp \vert^2}{ |q_f  B|} } \prj_+ \,
  { \slashed p_\parallel + m  \over p_\parallel^2- m^2}
  \Big|_{p^0\to p^0+i\varepsilon}
  \,, 
\\
\label{eq:Sar}
 S_{ar}(p) &=& 2 i\, e^{- \frac{ \vert \bm p_\perp \vert^2}{ |q_f  B|} } \prj_+ \,
  {\slashed p_\parallel+m \over p_\parallel^2- m^2}
  \Big|_{p^0\to p^0-i\varepsilon} 
\,, 
\\
\label{eq:Srr}
 S_{rr}(p) &=& \big[{1\over 2} - n_+(p^0)\big]\bigl[S_{ra}(p)  -S_{ar}(p)\bigr]
\, ,
\\
 S_{aa} (p) &=& 0
 \, ,
\end{eqnarray}
\end{subequations}
where $  n_\pm (p^0) = [\, e^{( p^0 \mp \mu)/T} + 1 \, ]^{-1}$ 
with a temperature $ T $ and chemical potential $ \mu $. 
Note that the Lorentz-boost invariance along constant magnetic fields 
is lost due to the presence of the medium. 
In this paper, we apply a constant magnetic field $ \bB $ 
in the medium rest frame where the fermion distribution function is given by $ n_\pm (p^0) $.

By the use of these propagators, the medium contribution to 
the retarded correlator is written down as 
\begin{eqnarray}
\Pi_R^{ \mu\nu} &=& \M_1^{ \mu\nu} + \M_2^{ \mu\nu} 
\, ,
\end{eqnarray}
where 
\begin{subequations}
\begin{eqnarray}
\label{eq:med1}
i \M_1^{ \mu\nu}  &=& - ( - iq_f)^2 \int \frac{d^4p}{(2\pi)^4} \tr[ \gam^\mu S_\ar(p+q) \gam^\nu S_\rr(p) ]
\, ,
\\  
\label{eq:med2}
i \M_2^{ \mu\nu}  &=& - (-i q_f)^2 \int \frac{d^4p}{(2\pi)^4} \tr[ \gam^\mu S_\rr(p+q) \gam^\nu S_\ra(p) ]
\, . 
\end{eqnarray}
\end{subequations} 
Note that one of the propagators $   S_{rr}(p) $ is proportional to the spectral density 
\begin{eqnarray}
\rho (p) = S_\ra(p) - S_\ar(p) 
= e^{-\frac{|\bp_\perp|^2}{ |q_f B| }} ( \slashed p_\parallel + m)   \prj_+
\cdot  \frac{2\pi}{\epsilon_p}  \left[ \delta( p^0  - \epsilon_p) - \delta( p^0  + \epsilon_p) \right]
\, ,
\end{eqnarray}
where $\epsilon_p = \sqrt{p_z^2+m^2}$. 
Having computed the vacuum part in the previous subsection, 
we drop the vacuum contributions that are independent of the temperature/chemical potential. 
Then, the medium-induced part in $   S_{rr}(p) $ reads 
\begin{eqnarray}
 S_{rr}(p) = - n_+(p^0)  e^{ - \frac{|\bp_\perp|^2}{ |q_f B| }} ( \slashed p_\parallel + m)   \prj_+
\cdot  \frac{2\pi}{\epsilon_p}  \left[ \delta( p^0  - \epsilon_p) - \delta( p^0  + \epsilon_p) \right]
\, .
\end{eqnarray}
One can perform the Dirac trace and the $ p^0 $ integral straightforwardly 
as described in Appendix~\ref{sec:comp-medium} in detail.

Notice that the tensor structure in Eq.~(\ref{eq:transverse-op}) exhausts 
all possible gauge-invariant structures composed of $ g_\para^{\mu\nu} $ and $ q_\para^\mu $ 
in the effective (1+1) dimensions.\footnote{
In the LLL contribution, the nonzero components of the polarization tensor $ \Pi_R^{ \mu\nu} $ 
construct a $2\times 2  $ matrix. 
There is only one independent degree of freedom after a reduction 
due to the symmetric property $ \Pi_R^{ \mu\nu} = \Pi_R^{ \nu\mu} $ and gauge invariance 
$ q_\mu \Pi_R^{ \mu\nu} =0$ with $ \nu = 0,3 $. 
There is no splitting to the longitudinal and transverse components with respect to the photon momentum. 
This counting is, however, not valid once the higher Landau levels contribute 
with additional tensors $ g_\perp^{\mu\nu} $ and $ q_\perp^\mu $ 
(see, e.g., Refs.~\cite{Hattori:2012je, Hattori:2017xoo}). 
} 
Therefore, the medium effect only modifies the scalar function in an additive form 
\begin{eqnarray}
\tilde  \Pi_\para (\omega, q_z) = \tilde  \Pi_\para^\vac(q_\para^2)  + \tilde  \Pi_\para ^\temp(\omega,q_z)
\, .
\end{eqnarray}
Breaking of the Lorentz-boost invariance makes 
the medium contribution $ \tilde  \Pi_\para ^\temp(\omega,q_z) $ 
depend on the photon energy $ \omega $ and momentum $ q_z $ separately, 
while the vacuum part $ \tilde  \Pi_\para^\vac(q_\para^2)  $ only depends on the scalar form $ q_\para^2 $. 
After straightforward but rather lengthy calculations summarized in Appendix~\ref{sec:comp-medium}, 
the medium contribution is obtained as [see Eq.~(\ref{eq:722}) in the appendix] 
\begin{eqnarray}
\tilde \Pi_\para^{\temp }(\omega, q_z) 
=  \pi  \, m^2 \!\! \int_{-\infty}^\infty \frac{d p_z}{2\pi \epsilon_p} 
\frac{ ( q_\parallel^2 + 2 q_z p_z) \,   [n_+(\epsilon_p)  + n_-(\epsilon_p)] } 
{  q_\parallel^2  ( p_z - \frac{1}{2} q_z) ^2 - \frac{\omega^2}{4 }  (q_\parallel^2 -4m^2) }
\label{eq:Pi_medium}
\, .
\end{eqnarray}
The denominator of the integrand has a definite sign when $0 < q_\para^2 < 4m^2 $, 
implying a smooth result of the integral. 
In contrast, when $q_\para^2 < 0$ or $ 4m^2 < q_\para^2 $, 
the denominator changes its sign as the integral variable $ p_z $ runs over the integral region. 
The two pole positions are found to be 
\begin{eqnarray}
p^{\pm}_z  := \frac{1}{2} \left( \, q_z \pm \omega \sqrt{1- 4m^2/q_\para^2 } \ \right) 
\, .
\end{eqnarray}
These observations imply that the integral has singularities depending on 
the photon momentum $ q_\para^2 $ 
and acquires an imaginary part as we inspect in the subsequent sections. 
While the integral form (\ref{eq:Pi_medium}) was shown in Ref.~\cite{Fukushima:2015wck}, 
it was not evaluated. 
The reader is also referred to related works 
for (1+1) dimensional QED \cite{Dolan:1973qd, Baier:1991gg, Kao:1998yt} .

\subsection{Birefringence from the resummed photon propagator}

In the previous subsections, we have found that the polarization tensor 
has only one gauge-invariant tensor structure 
\begin{eqnarray}
\label{eq:Pi_LLL}
\Pi_R^{\mu\nu} =   \Pi_\para  P_\para^{\mu\nu}  
\, ,
\end{eqnarray}
where $  \Pi_\para =  m_B^2 e^{-\frac{\vert \bq_\perp \vert^2}{2 \vert q_fB\vert}} \tilde  \Pi_\para $.  
When the photon momentum $  q_\para^2$ is comparable in magnitude to $  \Pi_\para  $, 
one needs to resum the self-energy correction to the photon propagator. 
Namely, we consider the resummation of the ring diagrams in between the free photon propagators 
\begin{eqnarray}
D_R^{\mu\nu}(q) \= D_0^{\mu\nu}(q) 
+ D_0^{\mu\a_1}(q) i \Pi_{R\a_1\b_1} D_0^{\b_1\nu}(q)
\nnb
&&
+ D_0^{\mu\a_1}(q) i \Pi_{R\a_1\b_1} D_0^{\b_1\a_2}(q)i  \Pi_{R\a_2\b_2} D_0^{\b_2\nu}(q)
+ \cdots
\, ,
\end{eqnarray}
where the free photon propagator is given as 
\begin{eqnarray}
i D_0^{\mu\nu}(q) =\frac{ 1 }{ q^2} P_0^{\mu\nu}  +  \xi_g \frac{q^\mu q^\nu}{ (q^2)^2}
\, ,
\end{eqnarray}
with $ P_0^{\mu\nu} = g^{\mu\nu} - q^\mu q^\nu/q^2 $ and the gauge parameter $  \xi_g $. 

The gauge-dependent term is not mixed with the polarization tensor at any order 
because of the transversality $ q_\mu  P_\para^{\mu\nu} = q_\mu  P_0^{\mu\nu} =0 $. 
Also, notice the associative properties 
$ P_{0\, \a}^{\mu} P_0^{\a \nu}  =   P_0^{\mu\nu} $, 
$ P_{\para\, \a}^{\mu} P_\para^{\a \nu}  =   P_\para^{\mu\nu} $, 
and $ P_{0\, \a}^{\mu} P_\para^{\a \nu} =  P_\para^{\mu\nu} $. 
Therefore, we obtain (see Refs.~\cite{Hattori:2012je, Hattori:2017xoo} for generalization\footnote{
The conventions here correspond to those in Ref.~\cite{Hattori:2012je} 
as $ \Pi_\para = - q_\para^2 \chi_1 $. 
Equation~(\ref{eq:Photon-prop-LLL}) is consistent with that in Eq.~(C.8) therein 
within the LLL approximation [see also Eq.~(7) therein for the conventions]. 
}) 
\begin{eqnarray}
\label{eq:Photon-prop-LLL}
i D_R^{\mu\nu}(q)
\= i D_0^{\mu\nu}(q) + i D_{0\, \a} ^{\mu} (q) P_\para^{\a\nu} 
\sum_{n=1}^\infty  \left( -  \frac{i}{q^2} \cdot i  \Pi_\para \right)^n 
\nnb
\=  \frac{1}{q^2} \left[ \, P_0^{\mu\nu} 
+ \frac{    \Pi_\para  }{ q^2  - \Pi_\para  } P_\para^{\mu\nu} 
+ \xi_g \frac{q^\mu q^\nu}{q^2}  \, \right] 
\, ,
\end{eqnarray} 
which serves as the inverse of the kinetic term such that 
\begin{eqnarray} 
 \left( q^2 \delta_\a^\mu - q^\mu  q_\a -  \Pi_{R\a}^\mu + \frac{1}{\xi_g}  q^\mu q_\a  \right)
i D_R^{\a\nu}(q) = g^{\mu\nu} 
\, .
\end{eqnarray} 
Especially, when an on-shell LLL current $ j^\mu_\LLL $ is coupled to the photon propagator, we have 
\begin{eqnarray}
j_{\LLL\, \mu} i  D_R^{\mu\nu}(q) 
=  \frac{    j_{\LLL }^\nu  }{ q^2 - \Pi_\para   } 
\, ,
\end{eqnarray}
where we used the Ward identity $ q_\mu j_{\LLL}^{ \mu}   =0$ 
and the fact that the LLL current $  j_{\LLL}^{ \mu}  $ is only nonvanishing in the $ 0,3 $ components.

%
%

The resummed photon propagator (\ref{eq:Photon-prop-LLL}) has poles at 
\begin{eqnarray}
\label{eq:poles}
 q^2=0 \quad {\rm and} \quad   q^2  -   \Pi_\para (q_\para;q_\perp) =0
 \,.
\end{eqnarray}
Clearly, the two polarization modes acquire the different dispersion relations. 
This can be understood intuitively as follows. 
The cyclotron orbits shrink in the strong magnetic field limit, 
and charged particles reside in quasi-one dimension along the magnetic fields.
Thus, only the electric fields oscillating in the magnetic-field direction 
can couple to the charged particles, and are modified by the polarization. 
The above results show that this picture works at the quantum level as well, 
and the LLL contributions to the polarization tensor only modify 
the dispersion relation of one of the two polarization modes.

Explicitly solving Eq.~(\ref{eq:poles}), one can get the refractive index $ n = |\bq|/\omega $ 
for each polarization mode. 
The occurrence of polarization-dependent refractive indices is called the {\it birefringence} 
in magnetic fields and especially the {\it vacuum birefringence} when there is no medium. 
When the polarization tensor becomes a complex-valued, 
so does the refractive index, implying damping of the photon flux in the parallel polarization mode. 
The polarization-dependent absorption is called the {\it dichroism}. 
The dichroism in the LLL acts like a polarizer that 
cuts off one of the polarization modes oscillating along the slit. 
We have already obtained the explicit form of the imaginary part in the vacuum contribution (\ref{eq:I-limits}) 
and also seen that the integral (\ref{eq:Pi_medium}) for the thermal contribution 
could have an imaginary part due to the singularity. 
We will obtain analytic results of the imaginary part in the next section 
and investigate how the imaginary part is modified by the medium contribution.

Modification of the dispersion relation is also important for off-shell photons 
that mediate interactions between the LLL fermions. 
The polarization tensor describes the screening effect on the force mediated by gauge bosons, 
which has been discussed in various contexts (see, e.g., 
Refs.~\cite{Gusynin:1998zq, Gusynin:1999pq, 
Fukushima:2011nu, Ozaki:2015sya, Fukushima:2015wck, Hattori:2016emy, 
Bandyopadhyay:2016fyd, Li:2016bbh, 
Hattori:2016lqx, Hattori:2016cnt, Hattori:2017xoo, Hattori:2017qio}). 
The screening property at the long spacetime scale is determined by 
the Debye screening mass. We will discuss it in Sec.~\ref{sec:screening}.

\section{Inspecting medium effects}

\label{sec:medium}

In this section, we evaluate the momentum integral in Eq.~(\ref{eq:Pi_medium}) 
and inspect the medium contribution to the polarization tensor. 
This provides us with basic and physical ideas about the medium modifications 
that are important for physical applications in the later sections. 


\subsection{Imaginary part of the polarization tensor: Thermally induced reactions}

Without the help of magnetic fields, real photons never decay in vacuum due to a simple kinematics; 
the massless on-shell condition is not compatible with 
finite values of the invariant mass in would-be final states of decay processes. 
However, magnetic fields supply a certain amount of momentum in the form of the Lorentz force, 
allowing decay of a single real photon to a fermion and antifermion pair. 
According to the cutting rule or the optical theorem, 
opening of such a kinematical window manifests itself in a nonzero imaginary part 
of the vacuum polarization tensor: 
\begin{eqnarray}
\label{eq:imag-vac}
\Im m [\tilde  \Pi_\para^\vac ] = -
\frac{2 \pi m^2   }{ \sqrt{ q_\para^2 (q_\para^2-4m^2) } }
 \, \sgn(\omega) \, \theta(q_\para^2 - 4m^2) 
\, ,
\end{eqnarray}
which originates from the imaginary part of the $ I $ function (\ref{eq:I}). 
The step function specifies the threshold condition, indicating that 
the photon energy should be larger than the pair mass for the decay process to occur. 
Notice that the threshold condition $ q_\para^2 \geq 4m^2 $ can be compatible with 
the massless on-shell condition $ q^2 = q_\para^2 + q_\perp^2 = 0 $ in the four dimensions. 

Here, we investigate the thermal contributions to the imaginary part. 
One can expect two-fold kinematical effects of the thermally populated fermions and antifermions as follows. 
Since some of the positive-energy states are occupied by thermal excitations, 
the Pauli-blocking effect should be manifest. 
Also, the inverse process, i.e., a pair annihilation to a single-photon, should occur 
with the fermion and antifermion pairs in the thermal medium. 
Keeping those points in mind, we shall closely look into the medium effects encoded in 
the imaginary part of the integral (\ref{eq:Pi_medium}).

With an infinitesimal imaginary part $ \omega \to \omega + i \epsilon $ for the retarded correlator, 
the integrand in Eq~(\ref{eq:Pi_medium}) has poles at  
\begin{eqnarray}
\label{eq:pair-momentum}
 p^{\pm}_z  = \frac{1}{2} \big( \, q_z \pm \omega \sqrt{1- 4m^2/q_\para^2 } \ \big) \pm i   \epsilon 
 \, .
\end{eqnarray}
Thus, the integrand in Eq.~(\ref{eq:Pi_medium}) can be rearranged 
in a simple and convenient form 
\begin{eqnarray}
\tilde \Pi_\para^{\temp }(\omega, q_z) 
=   \frac{ 2 \pi  m^2  }{ q_\para^2\sqrt{1- 4m^2/q_\para^2 }  }
 \int_{-\infty}^\infty \frac{d p_z}{2\pi \epsilon_p} 
[ n_+(\epsilon_p)  + n_-(\epsilon_p)]
\Big( \frac{ \epsilon_p^+ s_+ }{ p_z - p_z^+ } - \frac{ \epsilon_p^- s_-}{ p_z - p_z^- } \Big)
\label{eq:Pi_medium-1}
\, ,
\end{eqnarray}
where we defined 
\begin{subequations}
\label{eq:pair-energy0}
\begin{eqnarray}
\label{eq:energy-signs}
&&s_\pm = \sgn \Big(  \omega \pm q_z    \sqrt{1- 4m^2/q_\para^2 } \ \Big)
\, ,
\\
\label{eq:pair-energy}
&&\ep_p^\pm
= \frac{1}{2} s_\pm \Big (  \omega \pm  q_z   \sqrt{ 1 -  4m^2/q_\para^2 } \ \Big ) 
= \frac{1}{2} \big|   \omega \pm  q_z   \sqrt{ 1 -  4m^2/q_\para^2 } \, \big|
\, ,
\end{eqnarray}
\end{subequations}
when $  1 -  4m^2/q_\para^2 \geq 0 $. 
When $  1 -  4m^2/q_\para^2 \leq 0 $ and thus $ \ep_p^\pm $ is complex-valued, 
we promise that $ s_\pm =1 $ to maintain the original form 
\begin{eqnarray}
\label{eq:original}
\epsilon_p^\pm s_\pm 
= \frac{1}{2}  \Big (  \omega \pm   q_z   \sqrt{ 1 -  4m^2/q_\para^2 } \Big)
\, .
\end{eqnarray}
By the use of the familiar formula 
\begin{eqnarray}
\label{eq:formula1}
1/(x \pm i \epsilon) = P(1/x) \mp i\pi \delta (x) 
\, ,
\end{eqnarray}
one can extract the imaginary part of the polarization tensor as 
\begin{eqnarray}
\Im m \tilde   \Pi_\para^{\temp }
\=  \frac{ 2 \pi  m^2 } {   q_\parallel^2    \sqrt{ 1-4m^2/q_\para^2}  } 
\left[   \, s_+  \frac{n_+(\epsilon_p^+)  +  n_- (\epsilon_p^+)}{2}  
+  s_-   \frac{n_+(\epsilon_p^-)  +  n_- (\epsilon_p^-)}{2}   \, \right]
\label{eq:ImPi-med}
\, ,
\end{eqnarray} 
for $  1 -  4m^2/q_\para^2 \geq 0 $. 
We used the fact that 
\begin{eqnarray}
\ep_p^\pm = \sqrt{ (p^{\pm}_z)^2 + m^2 }
\, .
\end{eqnarray}

Next, we shall examine the sign function $s _\pm $ introduced just above. 
Since the polarization tensor is an even function of $ q_z  $, 
one can hereafter replace $ q_z $ by its absolute value. 
Note also that $ \omega^2 - (q_z  \sqrt{1 - 4m^2/q_\para^2})^2 = q_\para^2 [1 +  4m^2 (q_z/q_\para^2)^2] $, 
which shows a relative magnitude of the two terms in $ s_\pm $.  
The sign $  s_+$ is negative only when $  \omega <0$ and $ q_\para^2 > 4m^2 $, otherwise it is positive. 
Also, the sign $  s_-$ is positive only when $  \omega >0$ and $ q_\para^2 > 4m^2 $, otherwise it is negative. 
Summarizing the above cases, one finds that 
\begin{eqnarray}
\label{eq:signs}
s_\pm \=   \sgn(\omega) \,  \theta(q_\para^2 -4m^2) \pm \theta(-q_\para^2)
\, .
\end{eqnarray}
Plugging this expression into Eq.~(\ref{eq:ImPi-med}), 
one can reorganize the imaginary part as 
\begin{eqnarray}
\label{eq:med_final}
\Im m \tilde   \Pi_\para^{\temp }
=  \frac{ 2 \pi   m^2   }{  q_\parallel^2 \sqrt{ 1 -4m^2 /q_\parallel^2 } }  
\Bigg[   N_+   \sgn(\omega) \,  \theta( q_\para^2 - 4m^2)  +  N_-   \theta( - q_\para^2 )  \Bigg]
\, ,
\end{eqnarray}
where  
\begin{eqnarray}
N_\pm : =   \frac{n_+(\epsilon_p^+)  +  n_- (\epsilon_p^+)}{2}  
\pm   \frac{n_+(\epsilon_p^-)  +  n_- (\epsilon_p^-)}{2} 
\, .
\end{eqnarray}
Since $ N_+ $ ($ N_- $) is an even (odd) function of $ \omega $, 
the imaginary part $ \Im m  \Pi_\para^{\temp } $ is an odd function of $ \omega $.  
Adding the vacuum contribution (\ref{eq:imag-vac}) to the above, we obtain the total imaginary part 
\begin{eqnarray}
\label{eq:imag-sum}
\Im m \tilde   \Pi_\para \= \Im m\tilde   \Pi_\para^{\vac} + \Im m \tilde  \Pi_\para^{\temp} 
\\
\=  -  \frac{ 2 \pi   m^2   }{  q_\parallel^2 \sqrt{ 1 -4m^2 /q_\parallel^2 } }  
\Bigg[  (1- N_+)   \sgn(\omega) \,   \theta( q_\para^2 - 4m^2)  -  N_-   \theta( - q_\para^2 )  \Bigg]
\nn
\, .
\end{eqnarray}
Notice that a new kinematical window opens in the region $ q_\para^2 \leq  0 $, 
which was absent in vacuum.  
We examine the physical processes that give rise to 
the imaginary parts in the regions $ q_\para^2 > 4m^2 $ and $ q_\para^2 < 0 $, respectively. 
Below, we refer those regions ``time-like'' and ``space-like'' as in the (1+1)-dimensional case. 
The above result is symmetric under the sign flip of the chemical potential $ \mu \to - \mu $ as expected. 
The vacuum contribution should be maintained even at high-temperature and/or -density limit 
unlike in the hard thermal loop approximation in (3+1) dimensions; 
The temperature or density scale cannot appear as an overall factor on the dimensional ground. 
We will explicitly see interplay between the vacuum and thermal contributions in the following.

\subsubsection{Finite temperature}

We consider the zero-density case $ \mu = 0  $, 
where both the fermion and antifermion distribution functions reduce to the same form 
$ n_\pm (p^0) \to n(p^0) = [  e^{p^0/T} + 1 ]^{-1}  $. 
We first focus on the high-temperature limit such that $ T \gg \omega, q_z $. 
By the use of Eq.~(\ref{eq:signs}), we then obtain 
\begin{eqnarray}
N_+  \sim  1 - \frac{1}{4T}  ( \epsilon_p^+ + \epsilon_p^-)  =   1 - \frac{|\omega|}{4T} 
\, .
\end{eqnarray}
Observe that the first term in $ N_+ $ exactly cancels the vacuum contribution in Eq.~(\ref{eq:imag-sum}), i.e., 
\begin{eqnarray}
1- N_+  \sim      \frac{|\omega|}{4T} 
\, .
\end{eqnarray}
Consequently, the total imaginary part is highly suppressed 
in the time-like region by a factor of  $ |\omega|/T \ll 1$. 
This is due to the detail balance between the pair-creation and -annihilation processes. 
One can confirm this observation with a simple identity 
\begin{eqnarray}
\label{eq:N+---0}
1 - N_+  =    [ 1- n(\ep_p^+) ] [ 1-  n(\ep_p^-)]  - n(\ep_p^+)  n(\ep_p^-) 
\, .
\end{eqnarray}
The first term corresponds to the pair-creation channel $ \gam \to f \bar f $ 
that is subject to the Pauli-blocking effect in the final state, 
while the second term corresponds to the pair-annihilation channel $ f \bar f  \to  \gam $. 
According to the sign functions (\ref{eq:signs}), we confirm 
the energy conservation $ |\omega| =  \ep_p^+ + \ep_p^-$. 
The net pair creation occurs with a significant rate only in 
a sufficiently high-energy regime $\omega \gg T  $ where the occupation number is small 
or in the low-temperature limit $\omega , q_z \gg T  $ 
where the thermal contribution is exponentially suppressed. 
In such cases, the vacuum contribution stands as the dominant contribution 
to the imaginary part. 

On the other hand, one finds a new medium-induced channel in the space-like regime, 
that is, a contribution of the Landau damping to $N_-  $. 
The Landau damping is purely a medium effect, where a medium fermion (antifermion) 
is scattering off a space-like photon. 
In case of vacuum, the imaginary parts in the space-like regime 
cancel out in the final expression of the polarization tensor 
in all order of the Landau levels \cite{Hattori:2012je}. 
By the use of Eq.~(\ref{eq:signs}), we find that 
\begin{eqnarray}
- N_- \sim  \frac{1}{4T}  ( \epsilon_p^+ - \epsilon_p^-)  =  \frac{\omega}{4T} 
\, .
\end{eqnarray}
Similar to the time-like region, 
the imaginary part is again suppressed in the space-like region in the high-temperature limit. 
This is due to a detail balance as recognized in an identity 
\begin{eqnarray}
- N_- =    n(\ep_p^-)  [ 1- n(\ep_p^+) ] - n(\ep_p^+) [ 1-  n(\ep_p^-)] 
\, ,
\end{eqnarray} 
where the Pauli-blocking effect appears in the final states. 
The first (second) term corresponds to the gain (loss) term 
for the phase-space volume at $ p^0 = \ep_p^+ $. 
Namely, this identity shows the detail balance between the reciprocal channels 
 $ f + \gam^\ast \to f $ and $ f \to f  + \gam^\ast $. 
According to the sign functions (\ref{eq:signs}), we confirm 
the energy conservation $  \omega + \ep_p^- = \ep_p^+$.

 \begin{figure}[t]
\begin{minipage}{0.45\hsize} 
	\begin{center} 
\includegraphics[width=\hsize]{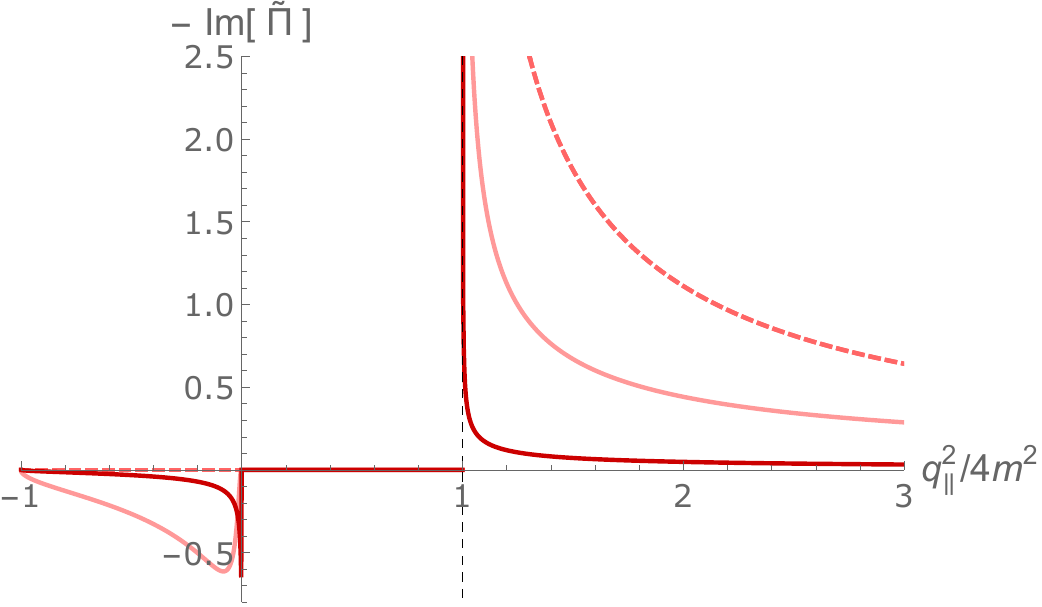}
	\end{center}
\end{minipage}
\begin{minipage}{0.45\hsize}
	\begin{center}
\includegraphics[width=\hsize]{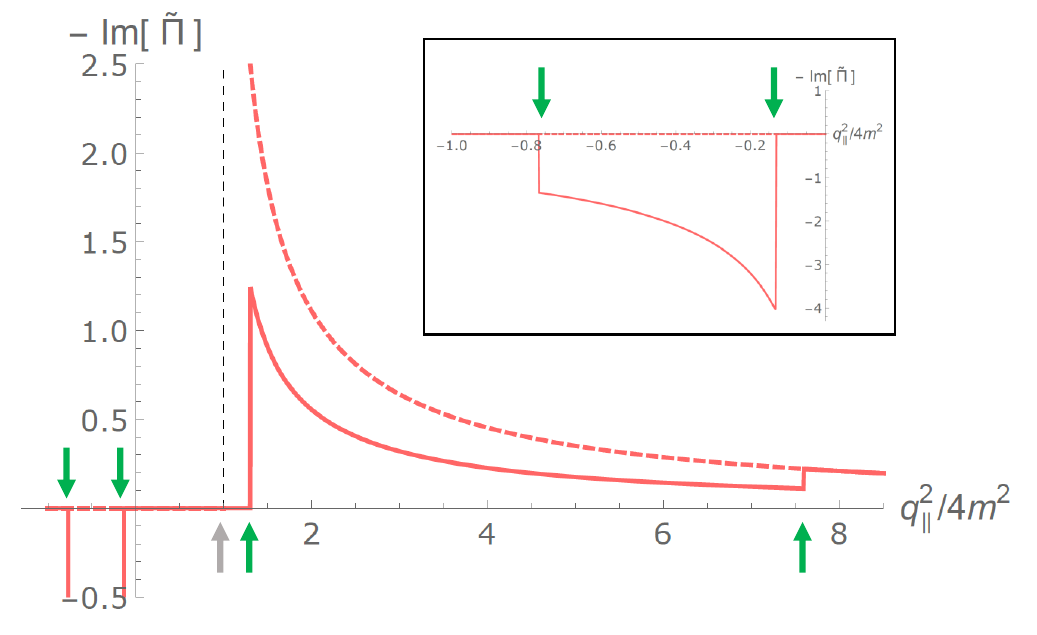}
	\end{center}
\end{minipage}
\caption{
The imaginary part of the polarization tensor at finite temperature (left) and density (right) 
as functions of $  \omega$. 
We take $ q_z/(2m) = 1$, $ T/(2m) =1, \, 10 $, and $ \mu/(2m) = 0 $ (left) 
and $ q_z/(2m) = 1$, $ T/(2m) =0 $, and $ \mu/(2m) = 1 $ (right). 
The solid lines include both the vacuum and medium contributions, 
while the dashed lines show the vacuum contribution alone ($ T=\mu=0 $). 
The step singularities are shown with the green arrows. 
The Landau-damping part is magnified in the in-cell window. 
}
\label{fig:imag-1}
\end{figure}

The imaginary part at finite temperature is plotted in the left panel of Fig.~\ref{fig:imag-1}, 
where we take $ q_z/(2m) = 1$, $ T/(2m) =1 , \, 10$ and $ \mu/(2m) = 0 $. 
The light and dark red lines are for $ T/(2m) =1 $ and $ 10$, respectively. 
The solid line includes both the vacuum and medium contributions, 
while the dashed line shows the vacuum contribution alone. 
We confirm the partial cancellation between the vacuum and medium contributions in the time-like regime 
and the occurrence of the Landau damping in the space-like regime.

Combining the expressions in the time-like and space-like regions, 
we obtain the high-temperature expansion of the total imaginary part  
\begin{eqnarray}
\label{eq:imag-total-expansion}
\Im m \tilde   \Pi_\para \sim -  \frac{  2 \pi  m^2  }{  q_\parallel^2 \sqrt{ 1 -4m^2 /q_\parallel^2 } }  
\cdot
\frac{\omega}{4T} \theta (1-  4m^2/q_\para^2 )
\, .
\end{eqnarray}
Also, the complete form at finite temperature is given by 
the sum of Eqs.~(\ref{eq:imag-vac}) and (\ref{eq:med_final}) as 
\begin{eqnarray}
\label{eq:imag-total}
\Im m \tilde   \Pi_\para  
=  -  \frac{ 2 \pi  m^2   }{  q_\parallel^2 \sqrt{ 1 -4m^2 /q_\parallel^2 } }  
\cdot
\frac{   \sinh( \frac{ \omega }{2T})  \, \theta (1-  4m^2/q_\para^2 )  }
{ \cosh( \frac{\omega}{2T}  ) + \cosh \big( \, \frac{q_z }{2T}  \sqrt{1-  4m^2/q_\para^2} \, \big)  }
\, .
\end{eqnarray}
Since all the terms in Eqs.~(\ref{eq:imag-vac}) and (\ref{eq:med_final}) are odd functions of $ \omega $, 
so is the final result in Eq.~(\ref{eq:imag-total}), 
which originates from a general property of the spectral density (see, e.g., Ref.~\cite{Kapusta:2006pm}).  
One can confirm that the zero-temperature limit of Eq.~(\ref{eq:imag-total}) 
agrees with the vacuum expression (\ref{eq:imag-vac}). 
It is useful to check the limit with the following expression 
\begin{eqnarray}
\label{eq:zero-T}
\lim_{T \to 0} \Im m \tilde   \Pi_\para  
= -   \frac{ 2  \pi  m^2   }{  q_\parallel^2 \sqrt{ 1 -4m^2 /q_\parallel^2 } }  
\lim_{T \to 0}   \frac{     \sgn(\omega) \,   \theta (1-  4m^2/q_\para^2 )  }
{ 1  + \exp \big [ \,  \frac{ 1 }{2T}  \big ( |q_z| \sqrt{1-  4m^2/q_\para^2} - |\omega| \big ) \, \big ]  }
\, .
\end{eqnarray} 
The limit is vanishing in the space-like region $ q_\para^2 < 0 $, 
while it reproduces the finite value of the imaginary part 
in vacuum in the time-like region $ q_\para^2 > 4m^2 $ [cf. Eq.~(\ref{eq:imag-vac})].

\subsubsection{Finite density}

\label{sec:imag-density}

At nonzero density $ (\mu \not = 0) $, 
the fermion and antifermion distribution functions should be distinguished from each other. 
Nevertheless, one can interpret the pair creation/annihilation and the Landau damping 
in a similar way to the zero-density case.

Similar to the finite-temperature case (\ref{eq:N+---0}), one finds an identity 
\begin{eqnarray}
1 - N_+ 
\= 1-  \frac12 \left[ \,  \{ n_+(\epsilon_p^+)  +  n_- (\epsilon_p^-)\}
+ \{  n_+(\epsilon_p^-)   +  n_- (\epsilon_p^+) \}   \, \right]
\nnb
 &=& \sum_{\a=\pm}  \frac12 [ \, \{ 1 - n_+(\ep_p^\a) \}\{ 1-  n_-(\ep_p^{-\a}) \} 
 -  n_+(\ep_p^\a)  n_-(\ep_p^{-\a})  \,]
 \label{eq:1-N+}
 \, .
\end{eqnarray}
This expression has the same form as in Eq.~(\ref{eq:N+---0}) 
up to the difference between the fermion and antifermion distribution functions. 
In the pair creation/annihilation channel, the fermion and antifermion can have different energies 
$ \ep_p^+ $ and $  \ep_p^- $ when the photon momentum is nonzero, i.e., $ q_z \not = 0 $. 
Thus, there are two kinematical windows depending on either a fermion or antifermion takes $ \ep_p^+ $. 
A finite chemical potential distinguishes those two cases, 
whereas they are degenerate at zero density (\ref{eq:N+---0}). 
Exchanging $ \ep_p^+ $ for $  \ep_p^- $ brings one kinematics to the other.

In the space-like region, the Landau damping of the fermions and antifermions occurs 
with different magnitudes. 
Therefore, separating the fermion and antifermion distribution functions, 
we find an identity 
\begin{eqnarray}
- N_- 
\= - \frac12 \left[ \,  \{ n_+(\epsilon_p^+)  -  n_+ (\epsilon_p^-) \}
+ \{  n_- (\epsilon_p^+ )   -  n_- (\epsilon_p^-) \}   \, \right]
\nnb
&=&  \sum_{c=\pm} \frac12 [ \,   n_{c}(\ep_p^-) \{ 1 - n_{c}(\ep_p^+) \}
 - n_{c}(\ep_p^+) \{1-  n_{c}(\ep_p^-) \}  \,]
\, .
\end{eqnarray}
The index $ c = + $ ($  - $) is for the Landau damping of fermions (antifermions), 
assuming that $ \mu > 0 $ without loosing generality.

At zero temperature, the distribution functions reduce 
to the step functions $ n_\pm(p^0) = \theta( \pm \mu - p^0 ) $. 
Thus, there is no annihilation channel, and only the pair-creation channel is left finite as 
\begin{eqnarray}
\label{eq:N+_T=0}
\left. 1 - N_+ \right|_{T=0} 
 &=& \frac12 [\,  \theta (\ep_p^+ -  \mu  )  + \theta ( \ep_p^- -  \mu  )  \,]
 \, .
\end{eqnarray}
Either fermion or antifermion is free of the Pauli-blocking effect 
since its positive-energy state is completely vacant. 
However, the other one needs to find an unoccupied state 
above the Fermi surface to avoid the Pauli-blocking, resulting in the above step functions. 
As for the Landau damping process, 
it occurs between an occupied initial state and an unoccupied final state. 
Thus, there are only fermion contributions (when we assume $ \mu > 0 $ as above). 
The gain and loss terms read  
\begin{eqnarray}
\label{eq:N-_T=0}
- \left.  N_- \right|_{T=0} 
&=&   \frac12 [ \,  \theta (  \mu - \ep_p^- ) \theta (\ep_p^+ -  \mu  ) 
-  \theta (  \mu - \ep_p^+ ) \theta (\ep_p^- -  \mu  )   \,]
\, .
\end{eqnarray} 
It is a simple task to confirm the zero-density limit $ (\mu \to 0) $, 
where we find that 
\begin{subequations}
\begin{eqnarray}
&&
\lim_{\mu \to 0} \left. 1 - N_+ \right|_{T=0} = 1
\, ,
\\
&&
\lim_{\mu \to 0} \left.  N_- \right|_{T=0} = 0
\, .
\end{eqnarray}
\end{subequations}
These limits reproduce the vacuum expression (\ref{eq:imag-vac}).

\begin{figure}[t]
\begin{minipage}{0.48\hsize} 
	\begin{center} 
\includegraphics[width=\hsize]{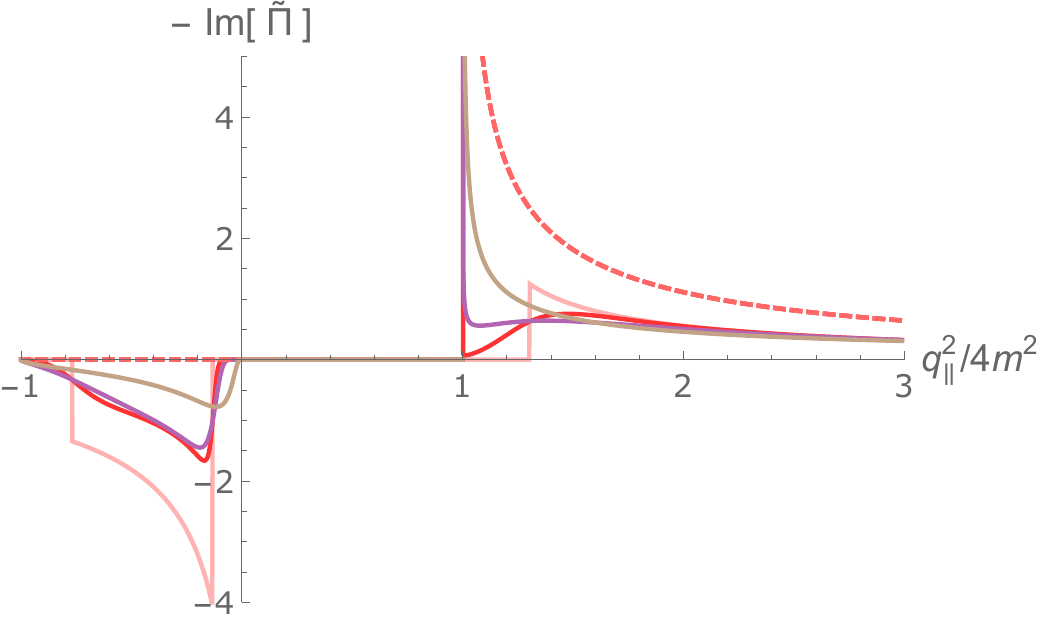}
	\end{center}
\end{minipage}
\begin{minipage}{0.48\hsize}
	\begin{center}
\includegraphics[width=\hsize]{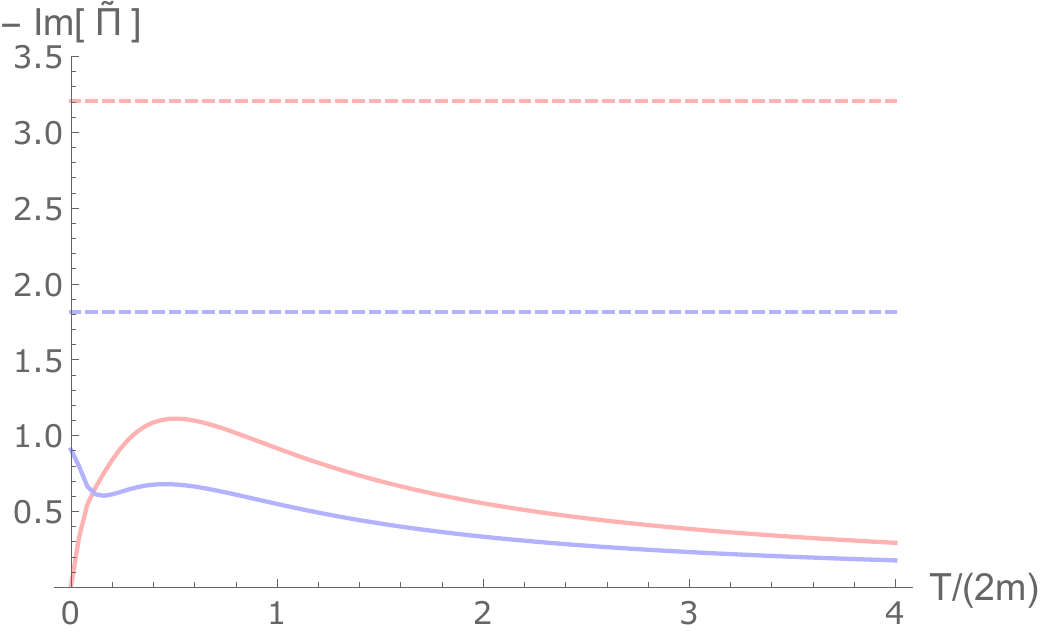}
	\end{center}
\end{minipage}
\caption{The imaginary part of the polarization tensor at finite temperature and density 
as functions of $  \omega$. 
Dashed lines show the vacuum contribution ($ T=\mu=0 $). 
We take $ q_z/(2m) = 1$, $ \mu/(2m) = 1 $ and $ T/\mu = \{ 0, \, 0.05 , \, 0.1 , \, 0.5 \}$ 
shown with the pink, red, purple and brown lines, respectively (left). 
Temperature dependences are shown at $ q_\para^2/(2m)^2 = 1.2 $ and $ 1.5 $ 
with the red and blue lines (right). The other parameters are the same as in the left panel. 
}
\label{fig:imag-2}
\end{figure}

The imaginary part at finite chemical potential is plotted in the right panel of Fig.~\ref{fig:imag-1}, 
where we take $ q_z/(2m) = 1$, $ T/(2m) =0 $ and $ \mu/(2m) = 1$.  
We confirm the occurrence of the Landau damping in the space-like regime. 
A negative peak appears in a sharp window due to the step singularities of 
the distribution functions at zero temperature (see also Fig.~\ref{fig:imag-2}). 
Remarkably, the threshold in the time-like regime is no longer located at $ q_\para^2 = 4m^2 $ (gray arrow) 
because the fermion states are occupied up to the Fermi surface. 
The new thresholds are specified by the conditions $ \epsilon_p^\pm > \mu $, 
requiring the threshold photon energy $ \omega_{\rm th}  
= \mu + \sqrt{q_z^2 + \mu^2 \pm 2|q_z| p_F \, }$ 
with the Fermi momentum $ p_F =  \sqrt{\mu^2-m^2} $ as indicated by the green arrows. 
There appear two thresholds from the two kinematics mentioned below Eq.~(\ref{eq:1-N+}), 
which clearly degenerate when $ q_z = 0 $. 
The shift and split of the thresholds are induced by the presence of a sharp Fermi surface. 
The singular threshold behavior in the vacuum contribution at $ q_\para^2 =4m^2 $ 
is completely cancelled out by the density effect at zero temperature. 
This exact cancellation is confirmed with an analytic expression in Appendix~\ref{sec:zero-T}. 
We also discuss the corresponding singular behaviors appearing in the real part 
in the next subsection.

Similar to the zero-density case~(\ref{eq:imag-total}), one can express the full form of 
the imaginary part with the hyperbolic functions. 
The complete form at finite temperature and density is found to be 
\begin{eqnarray}
\label{eq:imag-total-density}
\Im m \tilde   \Pi_\para  
\=  - \frac{ 2 \pi  m^2   }{  q_\parallel^2 \sqrt{ 1 -4m^2 /q_\parallel^2 }  }
\cdot \frac12 \sum_{c=\pm}
\frac{   \sinh( \frac{ \omega }{2T})  \, \theta (1-  4m^2/q_\para^2 )  }
{ \cosh( \frac{\omega}{2T}  ) + \cosh \big( \, \frac{q_z }{2T}  \sqrt{1-  4m^2/q_\para^2}
+ c \frac{\mu}{T} \, \big)  }
\, .
\end{eqnarray}
The chemical potential $ \mu $ appears only in the denominator. 
One can easily confirm that this expression reduces to Eq.~(\ref{eq:imag-total}) when $ \mu \to 0 $.

In Fig.~\ref{fig:imag-2}, we show the imaginary part of the polarization tensor 
at finite temperature and density. 
In the left panel, we take $ q_z/(2m) = 1$, $ \mu/(2m) = 1 $ and $ T/\mu = \{ 0, \, 0.05 , \, 0.1 , \, 0.5 \}$. 
This plot demonstrates the step singularity at zero temperature 
and its smearing at finite temperature; 
temperature effects smear the sharp edge of the Fermi surface 
and make vacancies near the Fermi surface. 
The threshold position in the time-like regime goes back to $ q_\para^2 = (2m)^2 $ 
as soon as the system gets an infinitesimal temperature as shown with the red line. 
Nevertheless, the imaginary part is still suppressed near the threshold 
since few vacancies are available near the smeared Fermi surface. 
The imaginary part in the time-like regime is lifted up as we further increase temperature. 
However, we expect that the imaginary part is finally suppressed at high temperature, 
since the antifermion states get occupied by the thermal excitations. 
Thus, the temperature dependence is not monotonic at finite density.

In the right panel of Fig.~\ref{fig:imag-2}, we show the temperature dependence 
at $ q_\para^2/(2m)^2 = 1.2 $ (red) and $ 1.5 $ (blue) on the both sides of the step singularity at $ T=0 $. 
The other parameters are the same as in the left panel. 
One confirms the above observations made in the left panel. 
Namely, the red line rises from zero as we increase temperature 
and hits a turnover point at a certain temperature. 
On the other hand, the blue line decreases starting from a finite value above the step-singularity threshold. 
The position of a local peak is roughly the same as that of the red line 
and shifts to higher temperature as we increase the chemical potential as expected. 
Overall, the imaginary part is suppressed by the medium effects.

Also, we confirm the step singularities (\ref{eq:N-_T=0}) in the space-like regime $ q_\para^2 < 0 $. 
The Landau damping emerges in a window where $ \epsilon_p^\mp < \mu < \epsilon_p^\pm  $ is satisfied. 
The step singularities are smeared by temperature effects that reduce the depth of the negative peak 
but create a tail toward larger $ |q_\para^2| $.

\subsection{Real part of the polarization tensor}

\label{sec:real_part}

In this section, we investigate the real part of the polarization tensor. 
The corresponding imaginary part is shown in Fig.~\ref{fig:imag-1}. 
According to the formula (\ref{eq:formula1}), the real part is given by Cauchy's principal value. 
Including the vacuum and thermal contributions in Eqs.~(\ref{eq:vac_massive}) and (\ref{eq:Pi_medium-1}), 
we have 
\begin{eqnarray}
\label{eq:real-total}
\Re e[ \tilde  \Pi_\para (\omega, q_z)] 
\= \Re e[ \, \tilde \Pi_\para^\vac  (q_\para^2) +  \tilde   \Pi_\para^\temp (\omega, q_z) \, ] 
\\
\=
 \Re e  \bigg[ \, 1-  I \big( \frac{q_\parallel^2}{4m^2} \big) 
+
m^2  \int_{-\infty}^\infty \frac{d p_z}{\epsilon_p} 
\frac{  n_+(\epsilon_p)  + n_-(\epsilon_p) }{ q_\para^2\sqrt{1- 4m^2/q_\para^2 }  }
\Big( \frac{ \epsilon_p^+ s_+ }{ p_z - p_z^+ } - \frac{ \epsilon_p^- s_-}{ p_z - p_z^- } \Big)
\,  \bigg] 
\nn
\, .
\end{eqnarray}
When $ 0 < q_\para^2 < 4m^2 $, the pole positions $ p^{\pm}_z $ take complex values 
\begin{subequations}
\begin{eqnarray}
&&
p^{\pm}_z  = \frac{1}{2} \left( \, q_z \pm i \omega \sqrt{ | 1- 4m^2/q_\para^2 | } \ \right) 
 \label{eq:p_z-complex}
 \, ,
\\
&&
 (p_z^\pm)^\ast = p_z^\mp
 \label{eq:p_z-conj}
\, .
\end{eqnarray}
\end{subequations} 
Thus, they are displaced from the real axis. 
On the other hand, the two poles $ p_z = p_z^\pm $ are located on the real axis 
in the kinematical regimes $ q_\para^2 < 0 $ or  $ 4m^2 < q_\para^2 $ 
where the on-shell conditions for the reactions, discussed in the previous subsections, are satisfied. 
The real part of the polarization tensor has resonant peaks near the boundaries 
between the kinematic regimes of the vanishing and nonvanishing imaginary parts. 
The correlation between the real and imaginary parts can be understood with the dispersion integral 
\begin{eqnarray}
\Re e [\tilde \Pi(\omega, q_z) ]
= \frac{1}{\pi} \int_{-\infty}^\infty d\omega' \frac{ \Im m [ \Pi(\omega',q_z) ] }{\omega' - \omega}
\, .
\end{eqnarray}
The integral value exhibits resonant peaks when the peak of the integral kernel at $ \omega' = \omega $ 
overlaps with the peaks of the imaginary part $  \Im m [ \Pi(\omega',q_z) ] $ 
that we discussed in the previous subsection.

\begin{figure}[t]
\begin{minipage}{0.48\hsize} 
	\begin{center} 
\includegraphics[width=\hsize]{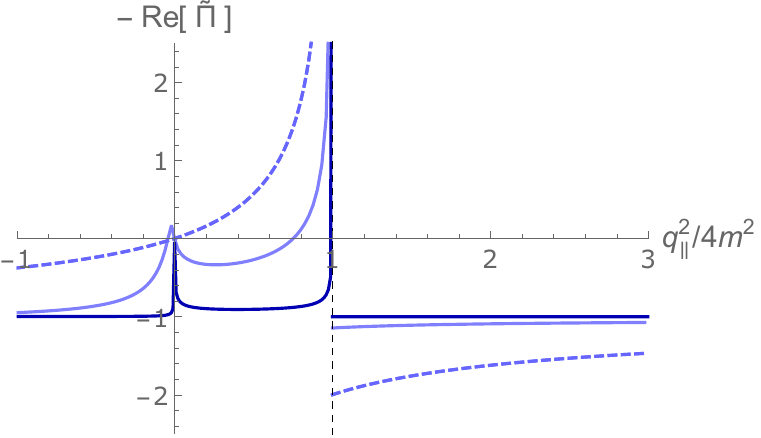}
	\end{center}
\end{minipage}
\begin{minipage}{0.48\hsize}
	\begin{center}
\includegraphics[width=\hsize]{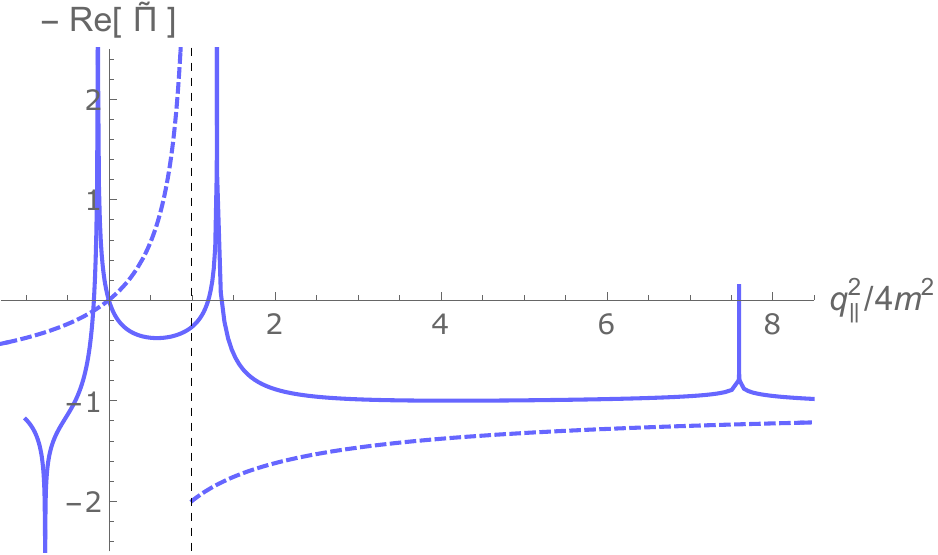}
	\end{center}
\end{minipage}
\caption{
The real part of the polarization tensor at finite temperature (left) and density (right) 
as functions of $  \omega$. 
We take the same parameters as in Fig.~\ref{fig:imag-1} for the imaginary part; 
$ q_z/(2m) = 1$, $ T/(2m) =1 $, $10 $, and $ \mu/(2m) = 0 $ (left) 
and $ q_z/(2m) = 1$, $ T/(2m) =0 $, and $ \mu/(2m) = 1 $ (right). 
The solid line include both the vacuum and medium contributions, 
while the dashed lines show the vacuum contribution alone ($ T=\mu=0 $). }
\label{fig:real-1}
\end{figure}

In Fig.~\ref{fig:real-1}, we show the numerical results for the real part. 
We take the same parameters as in Fig.~\ref{fig:imag-1} for the imaginary part. 
The sum of the vacuum and thermal contributions is shown by the solid line, 
while the vacuum contribution alone is shown by the dashed line. 
The left panel shows temperature effects at $ T/(2m) = 1 $ and $ 10 $ 
with the light and dark blue lines, respectively. 
As we increase temperature, the polarization tensor converges to 
the massless case (\ref{eq:Pi-massless}), i.e., the constant magnitude $ \Re e[\tilde \Pi _\para ] = 1 $. 
There are two resonant peaks associated with 
the Landau damping and the pair creation (cf. Fig.~\ref{fig:imag-1}). 
The singularities only go away in the infinite-temperature limit as shown explicitly 
in Eq.~(\ref{eq:Pi_infinite-T}) in an appendix. 
Note also that all the lines in the plots go through the origin at $ q_\para^2 = 0 $ 
as long as the temperature and/or density is finite. 
Namely, the integral for the thermal contribution vanishes in the limit $ q_\para^2 \to 0  $ 
with $ q_z \not = 0 $ since the vacuum parts vanish by themselves according to Eq.~(\ref{eq:I-limits}). 
When $ q_\para^2 \to 0  $ from below, both $  \ep_p^\pm $ and $  p^{\pm}_z  $ diverge, 
and we have $ s_\pm = \pm 1 $ and $  \ep_p^\pm /  p^{\pm}_z  \to \pm 1 $ 
according to the definitions~(\ref{eq:pair-momentum}) and (\ref{eq:pair-energy0}). 
Thus, the integral vanishes unless there were contributions from the asymptotic 
boundaries $ p_z \to \pm \infty $ that are indeed exponentially suppressed 
at finite temperature and/or density. 
When $ q_\para^2 \to 0  $ from above, 
both $  \ep_p^\pm $ and $  p^{\pm}_z  $ are complex-valued and their imaginary parts diverge. 
Then, we have 
\begin{eqnarray}
\label{eq:q2=0}
&& \hspace{-1cm}
\Re e  \int_{-\infty}^\infty \frac{d p_z}{\epsilon_p} 
\frac{  n_+(\epsilon_p)  + n_-(\epsilon_p) }{ q_\para^2\sqrt{1- 4m^2/q_\para^2 }  }
\Big( \frac{ \epsilon_p^+ s_+ }{ p_z - p_z^+ } - \frac{ \epsilon_p^- s_-}{ p_z - p_z^- } \Big)
\\
&\sim& 
 \int_{-\infty}^\infty \frac{d p_z}{\epsilon_p} 
\frac{  n_+(\epsilon_p)  + n_-(\epsilon_p) }{  q_\para^2\sqrt{|1- 4m^2/q_\para^2| }  }
\frac{ \Im m( \epsilon_p^+ p_z^- - \epsilon_p^- p_z^+ )  }{ |1- 4m^2/q_\para^2|  } 
\sim  -  \int_{-\infty}^\infty \frac{d p_z}{\epsilon_p} 
\frac{  n_+(\epsilon_p)  + n_-(\epsilon_p) }{  |1- 4m^2/q_\para^2|  }
\nn
\, ,
\end{eqnarray}
meaning that the integral value approaches zero from below 
when $ q_\para^2 \to 0 $ from above, as the plots consistently indicate.

The right panel shows density effects at zero temperature. 
One finds that there are four singular peaks, 
while the threshold singularity at $ q_\para^2 = 4m^2 $ is gone. 
The total polarization tensor is a smooth function at $ q_\para^2 = 4m^2 $, 
even though it contains the divergent vacuum contribution. 
It is thus expected that the singular threshold behavior in the $ I $ function (\ref{eq:I}) 
is exactly cancelled by the finite-density contribution, 
giving rise to new singularities at different positions. 
Indeed, we have found by inspecting the imaginary part in the previous subsection 
that the pair production threshold, originally located at $ q_\para^2 = 4m^2 $, 
is shifted and split into two thresholds due to the Pauli-blocking effect when $ q_z \not = 0 $.

To confirm the above characteristic behaviors at the zero temperature limit, 
we perform the integral (\ref{eq:real-total}) analytically. 
The Fermi-Dirac distribution functions reduce to the step functions that provide 
the integral with the cutoff at the Fermi momentum $ p_F = \sqrt{ \mu^2 - m^2} $ at zero temperature. 
Then, the integral can be performed as 
\begin{eqnarray}
\left.  \tilde  \Pi_\para^\temp (\omega, q_z) \right|_{T=0}
\= \frac{  m^2    }{ q_\para^2\sqrt{1- 4m^2/q_\para^2 }  }
\int_{-p_F}^{p_F} \frac{d p_z}{ \epsilon_p} 
\Big( \frac{ \epsilon_p^+ s_+ }{ p_z - p_z^+ } - \frac{ \epsilon_p^- s_-}{ p_z - p_z^- } \Big)
\nnb
\=  \frac{ m^2  }{ q_\para^2\sqrt{1- 4m^2/q_\para^2 }  } 
\Big[ 
s_+ \ln  \frac{ \mu  p_z^+ - p_F \, \epsilon_p^+ }  
{ \mu p_z^+ + p_F \, \epsilon_p^+  } 
-
s_- \ln  \frac{ \mu p_z^- - p_F \, \epsilon_p^- }  
{ \mu p_z^- + p_F \, \epsilon_p^-  } 
\Big] 
\label{eq:Re-zero-T-1}
\, .
\end{eqnarray}
This analytic result exhibits the aforementioned remarkable properties as we summarize below. 
The reader is referred to Appendix~\ref{sec:zero-T} for detailed analyses.

First, we examine the cancellation of the divergences when the photon momentum 
approaches $  q_\para^2 = 4m^2$ from below. 
When $ 1 - 4m^2 /q_\para^2 <0 $, that is, $ 0 < q_\para^2 < 4m^2 $, 
the medium contribution is a real-valued function 
\begin{eqnarray}
\left. \tilde \Pi_\para^{\temp }(\omega, q_z) \right|_{T=0}
\=  \frac{ 2 m^2  }{ q_\para^2\sqrt{ |1- 4m^2/q_\para^2| }  } 
\arg \Big(   \frac{ \mu  p_z^+ - p_F \, \epsilon_p^+ }  { \mu p_z^+ + p_F \, \epsilon_p^+  }  \Big)
\label{eq:Pi_zero-T-real-1}
\, ,
\end{eqnarray}
where ``$ \arg  $'' denotes the argument of the complex variable. 
We take the principal value within $ [-\pi,\pi] $. 
The argument in Eq.~(\ref{eq:Pi_zero-T-real-1}) evolves from 0 to $ +\pi $ as 
we increase the photon momentum from $  q_\para^2 =0 $ to $ 4m^2$. 
Therefore, the medium contribution (\ref{eq:Pi_zero-T-real-1}) diverges 
to positive infinity as $q_\para^2 \to 4m^2  $, 
which is a specific behavior at nonzero density and zero temperature. 
This divergence exactly cancels that from the vacuum contribution (\ref{eq:vac_massive}) 
that behaves as 
\begin{eqnarray}
\tilde  \Pi_\para^\vac  (q_\para^2) 
= -  \frac{2\pi m^2 }{ \sqrt{q_\para^2 (4m^2 - q_\para^2) } }
\, , 
\quad \quad q_\para^2 \to 4m^2
\label{eq:threshold-vac}
\, ,
\end{eqnarray} 
with $\arctan(\infty) = \pi/2  $. 
Since the divergence is associated with the threshold behavior of the pair creation, 
the absence of the divergence implies that the threshold is shifted to higher energies.

\begin{figure}[t]
\includegraphics[width=0.7\hsize]{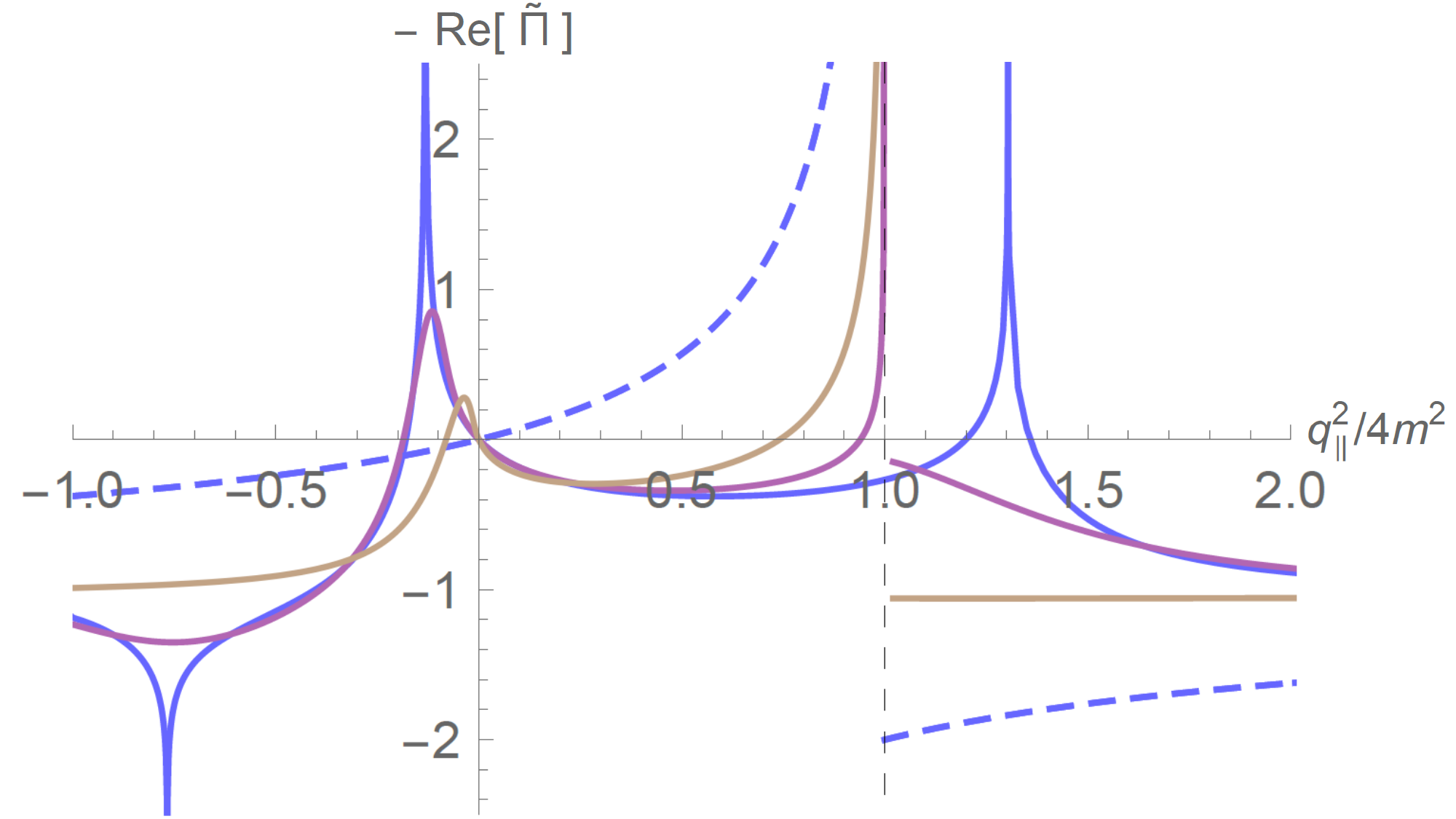}
\caption{The real part of the polarization tensor at finite temperature and density as functions of $  \omega$. 
We take the same parameters as in Fig.~\ref{fig:imag-2} for the imaginary part. 
Namely, we take $ q_z/(2m) = 1$, $ \mu/(2m) = 1 $ and $ T/\mu = \{ 0, \, 0.1 , \, 0.5 \}$ 
shown with the blue, purple and brown lines, respectively. 
The dashed line shows the vacuum contribution ($ T=\mu=0 $). 
}
\label{fig:real-2}
\end{figure}

When $ q_\para^2 \leq 0 $ or $4m^2 \leq q_\para^2 $, 
the arguments of the logarithms in Eq.~(\ref{eq:Re-zero-T-1}) are real-valued functions 
and can take both positive and negative values. 
When the arguments take negative values, the logarithms acquire imaginary parts, 
reproducing the previous result in Sec.~\ref{sec:imag-density}. 
The new threshold positions are given by $ \epsilon_p^\pm = \mu $. 
Accordingly, the real part exhibits divergences at the threshold positions 
where the arguments of the logarithms go through zeros. 
Therefore, we find that the singularities seen in the right panel of Fig.~\ref{fig:real-1} 
are logarithmic divergences. 
Note that the inverse square root, which gives the divergence at zero density 
or finite temperature, is regular at the new threshold positions.

In Fig.~\ref{fig:real-2}, we show temperature effects at the same chemical potential as above. 
The purple and brown lines show the finite-temperature results at $ T/\mu = 0.1$ and $ 0.5$ 
as in Fig.~\ref{fig:imag-2} for the imaginary part. 
The logarithmic divergences, observed at zero temperature, are smeared out 
and the divergent threshold behavior at $  q_\para^2=4m^2$ is 
instead restored with an infinitesimal temperature. 
Those behaviors reflect the fact that vacant states are available inside the Fermi surface 
at nonzero temperature (see the discussions about Fig.~\ref{fig:imag-2}). 
As we increase temperature, the line shape gets closer to 
those at zero density shown in the left panel of Fig.~\ref{fig:real-1}. 
As discussed around Eq.~(\ref{eq:q2=0}), all the lines go through the origin at $ q_\para^2=0 $.

\cout{

We perform the integral at the zero temperature limit. 
In the previous subsection, we have found that the pair production threshold, 
originally located at $ q_\para^2 = 4m^2 $, is shifted and split into two thresholds 
due to the Pauli-blocking effect. 
It is thus expected that the singular threshold behavior in the vacuum contribution $ I(q_\para^2/4m^2) $ 
is exactly cancelled by the thermal contribution, giving rise to two new singularities at different positions. 
The analytic result can provide an insight into such remarkable behaviors.

At zero temperature, the Fermi-Dirac distribution functions reduce to the step functions 
that provide the integral with the cutoff as  
\begin{eqnarray}
\label{eq:Re-zero-T}
\left.  \tilde  \Pi_\para^\temp (\omega, q_z) \right|_{T=0}
= \frac{  m_B^2 \, m^2    }{ q_\para^2\sqrt{1- 4m^2/q_\para^2 }  }
\int_{-p_F}^{p_F} \frac{d p_z}{ \epsilon_p} 
\Big( \frac{ \epsilon_p^+ s_+ }{ p_z - p_z^+ } - \frac{ \epsilon_p^- s_-}{ p_z - p_z^- } \Big)
\, ,
\end{eqnarray}
where $ p_F = \sqrt{ \mu^2 - m^2} $ is the Fermi momentum at zero temperature. 
Now, the integral can be performed by the use of an indefinite integral 
\begin{eqnarray}
 \int \frac{d p_z}{\epsilon_p} \frac{1}{p_z - a}
 = \frac{1}{ \sqrt{ a^2 + m^2}}
  \arctanh \left( - \frac{ a p_z + m^2 }{ \sqrt{ (p_z^2 + m^2) (a^2 +m^2) } } \right)
  + C
  \, ,
\end{eqnarray}
where $  C$ is the integral constant. 
Applying this formula, we obtain 
\begin{eqnarray}
\label{eq:integrals-zero-density}
 \int_{ -p_F}^{p_F} \frac{d p_z}{\epsilon_p} \frac{1}{p_z - p_z^\pm}
\= - \frac{1}{2 \epsilon_p^\pm} \Big[
\ln \left(  \frac{  \mu \epsilon_p^\pm + ( p_z^\pm p_F + m^2 ) } 
{  \mu \epsilon_p^\pm- ( p_z^\pm p_F  + m^2 ) } \right)
- 
\ln \left( \frac{  \mu \epsilon_p^\pm + ( - p_z^\pm p_F + m^2 ) } 
{  \mu \epsilon_p^\pm - ( - p_z^\pm  p_F  + m^2 ) } \right)
  \Big]
  \nnb
 \=  \frac{1}{ \epsilon_p^\pm} 
\ln \frac{ \mu p_z^\pm -  p_F \epsilon_p^\pm } {  \mu p_z^\pm + p_F  \epsilon_p^\pm  } 
 \, ,
\end{eqnarray}
where we used a relation, $\arctanh( z)   = - \frac12 \{ \ln (1- z) - \ln(1+z) \}  $. 
Plugging this result into Eq.~(\ref{eq:Re-zero-T}), we arrive at a simple form 
\begin{eqnarray}
\left. \tilde \Pi_\para^{\temp }(\omega, q_z) \right|_{T=0}
\=  \frac{ m^2 \,  m_B^2 }{ q_\para^2\sqrt{1- 4m^2/q_\para^2 }  } 
\Big[ 
s_+ \ln  \frac{ \mu  p_z^+ - p_F \, \epsilon_p^+ }  
{ \mu p_z^+ + p_F \, \epsilon_p^+  } 
-
s_- \ln  \frac{ \mu p_z^- - p_F \, \epsilon_p^- }  
{ \mu p_z^- + p_F \, \epsilon_p^-  } 
\Big]
\label{eq:Pi_zero-T}
\, .
\end{eqnarray}
When $ 1 - 4m^2 /q_\para^2 <0 $, we have complex-conjugate properties 
$ (\epsilon_p^\pm)^\ast =  \epsilon_p^\mp $ and $ (p_z^\pm)^\ast = p_z^\mp $. 
Remember also that $ s_\pm=1 $ when $ 1 - 4m^2 /q_\para^2 <0 $ 
as promised above Eq.~(\ref{eq:original}). 
Thus, in this region, the medium contribution is a real-valued function 
\begin{eqnarray}
\left. \tilde \Pi_\para^{\temp }(\omega, q_z) \right|_{T=0}
\=  \frac{ 2 m^2 \,  m_B^2 }{ q_\para^2\sqrt{ |1- 4m^2/q_\para^2| }  } 
\arg \Big(   \frac{ \mu  p_z^+ - p_F \, \epsilon_p^+ }  { \mu p_z^+ + p_F \, \epsilon_p^+  }  \Big)
\label{eq:Pi_zero-T-real-1}
\, ,
\end{eqnarray}
where ``$ \arg  $'' denotes the argument of the complex variable. 
We take the principal value within $ [-\pi,\pi] $. 

}

\section{Medium-induced photon masses}

\label{sec:screening}

Having obtained the polarization tensor, we now investigate how 
it modifies the photon properties in strong magnetic fields. 
Specifically, we compute the photon masses that characterize 
one of the most fundamental properties of photons. 
According to Eq.~(\ref{eq:Photon-prop-LLL}), the pole position of the parallel mode is given as 
\begin{eqnarray}
\label{eq:pole}
q^2 -  \Pi_\para (q_\para; q_\perp) = 0
\, .
\end{eqnarray}
The other mode is not modified by the polarization of the LLL fermions as already discussed there.

In the massless limit, the polarization tensor reduces to the constant form (\ref{eq:Pi-massless}). 
Thus, the photon dispersion relation is immediately obtained as 
\begin{eqnarray}
\label{eq:massless-dispersion}
\omega^2 = |\bq|^2 + m_B^2 \, e^{-\frac{\vert \bq_\perp \vert^2}{2 \vert q_fB\vert}} 
\, .
\end{eqnarray} 
This is an analogue of the Schwinger mass \cite{Schwinger:1962tn, Schwinger:1962tp} 
that appeared here because of the effective dimensional reduction to the (1+1) dimensions 
along the strong magnetic fields.

Computing the photon dispersion relation with massive fermions is more involved due to 
the momentum dependence of the polarization tensor; 
one needs to solve Eq.~(\ref{eq:pole}) with respect to the frequency. 
In general, the polarization tensor exhibits different limiting behaviors 
in the vanishing frequency and momentum limits; those limits do not commute with each other. 
Thus, there are two different photon masses (see, e.g., Ref.~\cite{Bellac:2011kqa}). 
They are called the Debye screening mass and the plasma frequency, 
which we discuss below in order.

First, we consider a static (heavy) charge that is screened by the polarization effect. 
In the static limit, the pole of the photon propagator is given as $   |\bq|^2 + \Pi_\para (\omega=0) = 0$, 
and the polarization tensor cuts off the long-range propagation of a spatial photon. 
This is the ``Debye screening mass'' defined as 
\begin{eqnarray}
m_\Db^2 \=  \lim_{q_z/m \to 0}  \Pi_\para( \omega=0 , q_z; q_\perp=0) 
\nnb
\=  \lim_{q_z/m \to 0}  \Pi_\para^{\temp }( \omega=0  , q_z; q_\perp=0) 
\end{eqnarray}
We have first taken the vanishing frequency limit 
and the vanishing transverse-momentum limit $(  q_\perp=0)$. 
In the present case, the ordering of the limits does not matter for $ q_\perp $. 
The second equality means that there is no vacuum contribution to $ m_\Db^2 $ 
as long as the fermion mass $m  $ is finite, 
according to the property of the $  I$ function in Eq.~(\ref{eq:I-limits}); 
Massive fermions are not excited by photons with an infinitesimal energy. 
Inserting the medium contribution (\ref{eq:Pi_medium-1}), we have 
\begin{eqnarray}
m_\Db^2 
\=  -  m_B^2   \lim_{\bar q_z \to 0}   \frac{1}{ 2 \bar q_z} 
 \prj \int_{-\infty}^\infty \frac{d \bar p_z}{ \bar \epsilon_p} 
\frac{ n_+(m \bar \epsilon_p)  + n_-(m \bar \epsilon_p) } { \bar p_z - \bar q_z }
\label{eq:Debye-sc}
\, .
\end{eqnarray}
We introduced a dimensionless variable $ \bar q_z = q_z/(2m) $. 
The integral variable was scaled as $ p_z \to p_z' = p_z/m $ accordingly. 
For later use, other variables are also normalized as 
$ \bar \epsilon_p = \epsilon_p/m $, $ \bar T = T/m $, and $ \bar \mu = \mu/m $. 
The integral should be understood as the Cauchy principal value denoted with $ \prj $. 
For a small $ \bar q_z $, the integral value should be of order $ \bar q_z^1 $ 
because the integrand is an odd function with respect to $ \bar p_z $ when $ \bar q_z \to 0 $. 
Therefore, the Debye mass is determined by the first derivative of 
the integral value with respect to $  \bar q_z$. 
The integral value is dominated by the infrared contribution such that $ \bar p_z \sim \bar q_z \to 0 $ 
due to the effective dimensional reduction to the (1+1) dimensions. 
We will evaluate the integral in the next subsection. 

%

Next, we consider the dispersion relation of an on-shell photon, 
which can be regarded as a collective excitation composed of a photon and oscillating plasma. 
Photon dispersion relations may be no longer gapless in the absence of the Lorentz symmetry. 
The finite energy gap is the ``plasma frequency'' induced by the medium response to the photon field. 
To find the magnitude of the energy gap at $ |\bq| =0 $, one should take the vanishing momentum limit first. 
Therefore, the plasma frequency $  \omega_p$ is determined by the following equation 
\begin{eqnarray}
\omega_p^2 =  \Pi_\para(\omega_p, 0) 
=  m_B^2   \bigg[ \, 1- I \big( \frac{\omega_p^2}{4m^2} \big)
+ 2  m^2 \prj \int_{-\infty}^\infty \frac{d p_z}{ \epsilon_p} 
\frac{  n_+(\epsilon_p)  + n_-(\epsilon_p) } {   (2 \epsilon_p)^2 -  \omega_p^2  }
 \,  \bigg] 
 \, .
\end{eqnarray}
The right-hand side is still a function of the frequency $ \omega_p $, 
so that one needs to solve the above equation explicitly. 
Note that the first two terms in the brackets come from the vacuum contribution. 
One needs to maintain those terms in addition to the medium contribution, 
because the medium contributions does not necessarily dominate over the vacuum contribution 
due to the effective dimensional reduction. 
This contrasts to the four dimensional case where  the medium contribution, which is 
proportional to $ (q_f T)^2 $ or $ (q_f \mu)^2 $, governs the polarization effects 
at the high temperature/density regime. 
We will see shortly that the vacuum contribution plays a crucial role in the present case.

We again introduce dimensionless variables $ \bar \omega_p = \omega_p /(2m) $ and 
$ \bar m_B = m_B/(2m) $ to find 
\begin{eqnarray}
\label{eq:plasma-frequency}
\frac{ \bar \omega_p^2 }{ \bar m_B^2 }
=  1- I ( \bar \omega_p^2)
+ \frac12 \prj  \int_{-\infty}^\infty \frac{d \bar p_z}{  \bar \epsilon_p }
\frac{  n_+( m  \bar \epsilon_p  )  + n_-( m  \bar \epsilon_p) } {   \bar \epsilon_p^2 - \bar \omega_p^2  }
 \, .
\end{eqnarray}
Notice that the medium parameters, $ T $ and $\mu $, 
and the magnetic-field strength, solely encoded in $ \bar m_B $, 
are separated from each other on the different sides of the equation. 
One can find the solutions for general parameter sets 
by numerically evaluating the integral on the right-hand side of Eq.~(\ref{eq:plasma-frequency}) 
and then finding the intersection of the both sides as functions of $ \bar \omega_p^2 $.  

In the following, we discuss the Debye mass and the plasma frequency 
with both analytic and numerical methods. 

\subsection{Debye screening mass}

To understand the basic behavior of the Debye screening mass, 
we first consider the high temperature and/or chemical potential limit. 
In such cases, the fermion distribution functions provide a cutoff scale at 
temperature $ T $ or chemical potential $ \mu $. 
On the other hand, the remaining part of the integrand is convergent in a large $ p_z $ region by itself. 
Thus, in high-$T  $ and/or -$ \mu $ limit, the distribution functions can be simply replaced 
by the upper and lower boundaries of the integral that can be simply performed as 
\begin{eqnarray}
m_\Db^2 
= - m_B^2   \lim_{\Lambda/m\to \infty }   \lim_{ \bar q_z \to 0}   \frac{1}{ 2 \bar q_z} 
\prj \int_{- \Lambda/m}^{\Lambda/m} \frac{d \bar p_z}{ \bar \epsilon_p} 
\frac{ 1 } { \bar p_z - \bar q_z }
= m_B^2
\label{eq:Debye-sc-infinite}
\, .
\end{eqnarray}
The cut-off $ \Lambda $ is given by a large value of $ T $ or $ \mu $. 
In this limit, the Debye mass agrees with the Schwinger mass. 
Note that a hierarchy $ T/\mu \gg1 $ ($ T/\mu \ll 1 $) needs to be satisfied for the above replacement 
in addition to the high-temperature limit $\bar T \to \infty  $ (high-density limit $ \bar \mu \to \infty$).

\begin{figure}
\begin{minipage}{0.45\hsize} 
	\begin{center} 
\includegraphics[width=\hsize]{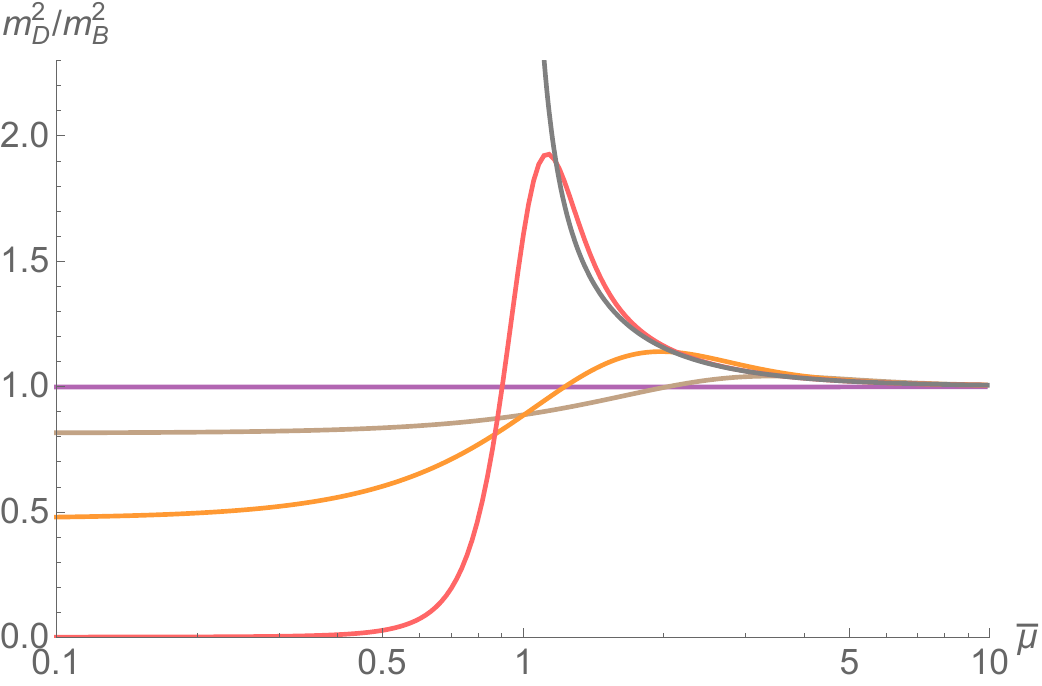}
	\end{center}
\end{minipage}
\begin{minipage}{0.45\hsize}
	\begin{center}
\includegraphics[width=\hsize]{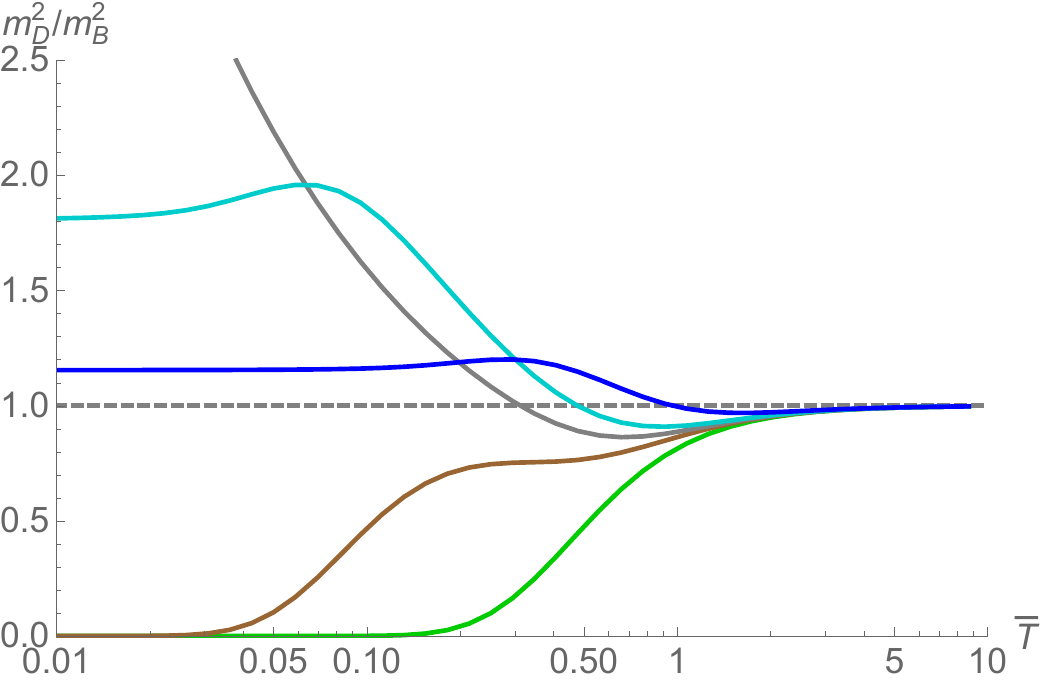}
	\end{center}
\end{minipage}
\caption{The Debye mass versus chemical potential (left) and temperature (right). 
The other parameter is taken as $ \bar T = \{0, \, 0.1, \, 0.5, \, 1, \, 10 \} $ and 
$\bar  \mu = \{ 0, \, 0.8, \, 1,\, 1.2, \, 2 \} $, respectively. 
In each plot, the gray line shows the case of $ \bar T = 0 $ and $ \bar \mu = 1 $ respectively. 
}
\label{fig:Debye}
\end{figure}

Next, the integral can be still performed analytically at zero temperature as 
\begin{eqnarray}
m_\Db^2 =  - m_B^2   \lim_{\bar q_z \to 0}   \frac{1}{ 2 \bar q_z} 
\prj \int_{-\sqrt{\bar \mu^2 -1}}^{\sqrt{\bar \mu^2 -1}} \frac{d \bar p_z}{ \bar \epsilon_p} 
\frac{ 1} { \bar p_z - \bar q_z }
= m_B^2 \,  \frac{  \bar \mu  }{ \sqrt{\bar \mu^2-1} }  
\label{eq:Debye-sc-zero-T}
\, ,
\end{eqnarray}
for $\bar  \mu > 1 $. 
The integral vanishes when $\bar  \mu\leq1 $, and so does the Debye mass 
\begin{eqnarray}
m_\Db^2 =0 
\, .
\end{eqnarray} 
This is because there is no fermion excitation at strict zero temperature when $\bar  \mu\leq1 $. 
The Debye mass at zero temperature is shown by the gray line in the left panel of Fig.~\ref{fig:Debye}. 
The Debye mass increases as we decrease $ \bar \mu $ from above 
and diverges at $ \bar \mu = 1 $; there is a discontinuity at $ \bar \mu = 1 $. 
To understand the diverging behavior, remember that the Debye mass (\ref{eq:Debye-sc}) is 
given by the first derivative of the integral value with respect to $ \bar q_z $. 
The factor of $ 1/(\bar p_z - \bar q_z) $ has a (diverging) peak near $ \bar p_z = 0 $ 
when $ \bar q_z $ is infinitesimally small. 
Thus, the integral value has a larger sensitivity to $ \bar q_z $ 
when the boundaries of the integral region at the Fermi momentum $ \bar p_z = \pm \sqrt{\bar \mu^2 -1} $ 
is located on a steeper point of the peak in the infrared region. 
This behavior is specific to the (1+1)-dimensional system, 
originating from the infrared behavior of the correlator. 
It contrasts to the well-known form of the Debye mass $ m_\Db^2 \sim ( q_f \mu)^2 $ 
from the hard dense loop in the four dimensions \cite{Bellac:2011kqa}, 
where the factor of $ \mu^2 $ originates from the integral value dominated by the hard momentum region.

One needs to perform numerical integration at finite temperature. 
The singularity at zero temperature is smeared by even an infinitesimal temperature. 
Nevertheless, one finds a finite-height peak structure at $ \bar T= 0.1 $ 
shown by the red line in Fig.~\ref{fig:Debye}. 
As we decrease temperature, the peak becomes higher and narrower, 
reproducing the zero-temperature behaviors discussed above. 
On the other hand, as we increase temperature, 
the Debye mass monotonically increases in the dilute region $\bar  \mu\ll1 $. 
Here, we take $ \bar T = \{0,\,  0.1, \, 0.5, \, 1, \, 10 \} $ for 
the lines shown in \{gray, red, orange, brown, purple\}. 
The behavior near $ \bar \mu = 1 $ can be understood as a smeared peak 
originating from the divergence at zero temperature, though it is rather involved. 
Finally, the Debye mass approaches the Schwinger mass when $\bar  \mu\gg1 $. 
Overall, when we increase temperature, 
the Debye mass increases (decreases) when $\bar \mu <1  $ ($\bar \mu > 1  $), 
exhibiting the opposite tendencies divided by $\bar \mu = 1  $.

The opposite tendencies can be understood from the density of occupied states 
in the low-momentum regime that dominantly contributes to the integral (\ref{eq:Debye-sc}). 
When $\bar \mu <1  $, the density of occupied low-momentum states increases from zero 
as we increase temperature, enhancing the Debye mass. 
Contrary, when $\bar \mu > 1  $, the occupied states inside the Fermi sphere 
are boiled up to the higher-momentum states, 
leaving fewer occupied states in the low-momentum regime behind. 
The Debye mass is thus suppressed as we increase temperature.

This opposite tendencies manifest themselves in the $ \bar T $ dependencies 
shown in the right panel of Fig.~\ref{fig:Debye}. 
The Debye mass approaches the Schwinger mass when $ \bar T \gg1  $ 
for all the shown values of $  \bar \mu$. 
As we decrease temperature, the lines splits near $ \bar T =1 $ depending on the value of $ \bar \mu $. 
We take $\bar  \mu = \{ 0, \, 0.8, \, 1,\, 1.2, \, 2 \} $ for the lines shown in 
\{green, brown, gray, blue, cyan\}. 
The Debye mass approaches zero when $\bar \mu <1  $ due to the absence of the excitation, 
while it increases when $\bar \mu \geq 1  $ induced by a larger density of 
the occupied states near the Fermi surface. 
These tendencies are consistent with those observed in the left panel. 
The heights of the plateaus in the small $ \bar T $ region correspond to the gray line in the left panel. 
The diverging behavior at $ \bar \mu =1 $, shown by the gray line in the right panel, 
corresponds to the divergence at $ \bar T = 0 $ show by the gray line in the left panel.

Summarizing, the Debye screening mass is dominantly induced by 
the fermions in the low-energy regime, and is enhanced when those states are occupied. 
Notice also that the Debye mass is scaled by the Schwinger mass $ m_B^2 $. 
The Debye mass only depends on the magnetic-field strength 
through the Landau degeneracy factor $|q_fB|/(2\pi)  $ in $ m_B^2 $. 
The density of degenerate states increases as we increase the magnetic-field strength, 
and thus naturally enhances the Debye mass.

\subsection{Plasma frequency}

Next, we investigate the plasma frequency that is an energy gap in the photon dispersion relation. 
It is obtained as the solutions of Eq.~(\ref{eq:plasma-frequency}).

\subsubsection{Vacuum case}

We first confirm that the plasma frequency vanishes in vacuum, i.e., 
\begin{eqnarray}
\omega_p = 0 
\, .
\end{eqnarray}
This is expected just because there are no on-shell fermions. 
Technically, this result originates from the behavior of the first two terms in Eq.~(\ref{eq:plasma-frequency}). 
One can show that 
$ 1- I ( \bar \omega_p^2) = - \frac23  \bar \omega_p^2 + \order( \omega_p^4) \to 0$ 
as $ \bar \omega_p^2 \to 0 $ 
and that this function monotonically decreases to 
negative infinity $  1- I ( \bar \omega_p^2) \to - \infty $ 
when $ \bar \omega_p^2  \to 1$ from below. 
This behavior is confirmed with the numerical plot in the left panel of Fig.~\ref{fig:PF-temp} 
shown by the gray solid line. 
Therefore, there is no intersection with the positive slope 
(shown by the blue line in Fig.~\ref{fig:PF-temp}) except for the one at the origin, 
meaning that the photon dispersion relation remains gapless in magnetic fields.

\subsubsection{Finite temperature}

We now include the integral from the medium contribution, i.e., 
the last term in Eq.~(\ref{eq:plasma-frequency}). 
This integral is positive definite at $ \omega_p^2 = 0 $. 
Therefore, the right-hand side of Eq.~(\ref{eq:plasma-frequency}) 
is lifted up from zero to a positive value at $ \omega_p^2 = 0 $, 
making a crucial difference from the vacuum case.

The limiting behavior near the threshold $ \bar \omega_p^2  \to 1$, when approached from below, 
needs to be examined carefully due to the fermion mass dependence. 
In the infinite-temperature limit $ \bar T = T/m \to \infty $ and $\bar T /\bar \omega_p^2  \to \infty$, 
one can replace the fermion distribution functions by the limiting value $  n_\pm \to 1/2$, 
and perform the integral as 
\begin{eqnarray}
\label{eq:integral-infinite-T}
 \frac12  \int_{-\infty}^\infty \frac{d \bar p_z}{  \bar \epsilon_p }
\frac{  n_+( m  \bar \epsilon_p  )  + n_-( m  \bar \epsilon_p) } {   \bar \epsilon_p^2 - \bar \omega_p^2  }
\to \frac12 \int_{-\infty}^\infty \frac{d \bar p_z}{  \bar \epsilon_p }
\frac{  1 } {  \bar \epsilon_p^2 - \bar \omega_p^2  }
=  I ( \bar \omega_p^2) 
\, .
\end{eqnarray}
Notice that the integral agrees with the $ I   $ function, 
and exactly cancels the same function from the vacuum contribution in Eq.~(\ref{eq:plasma-frequency}); 
The massless limit is reproduced. 
This means that the plasma frequency is given by the Schwinger mass 
\begin{eqnarray}
\omega_p = m_B
\, .
\end{eqnarray}
Especially, it is remarkable that the divergences from the vacuum and medium contributions 
have been exactly cancelled with each other.

\begin{figure}
\begin{minipage}{0.45\hsize} 
	\begin{center} 
\includegraphics[width=\hsize]{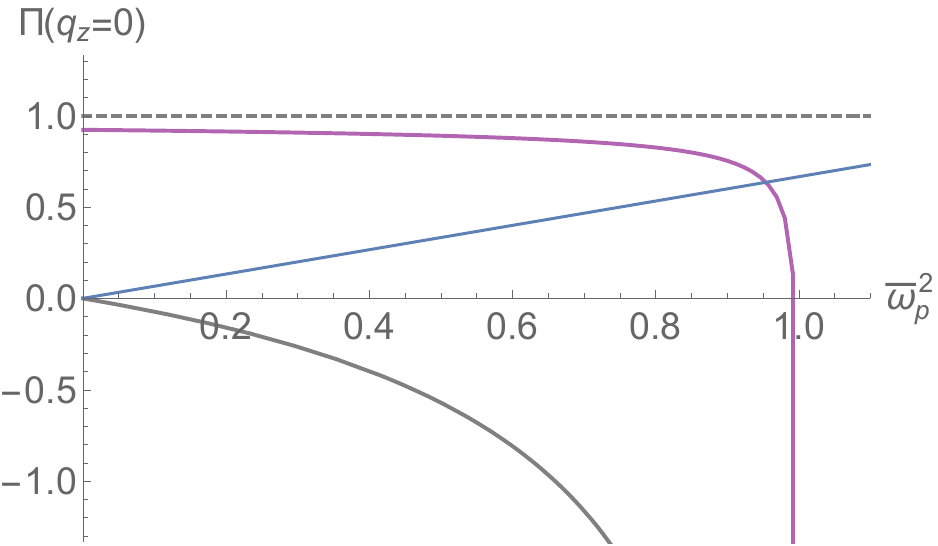}
	\end{center}
\end{minipage}
\begin{minipage}{0.45\hsize}
	\begin{center}
\includegraphics[width=\hsize]{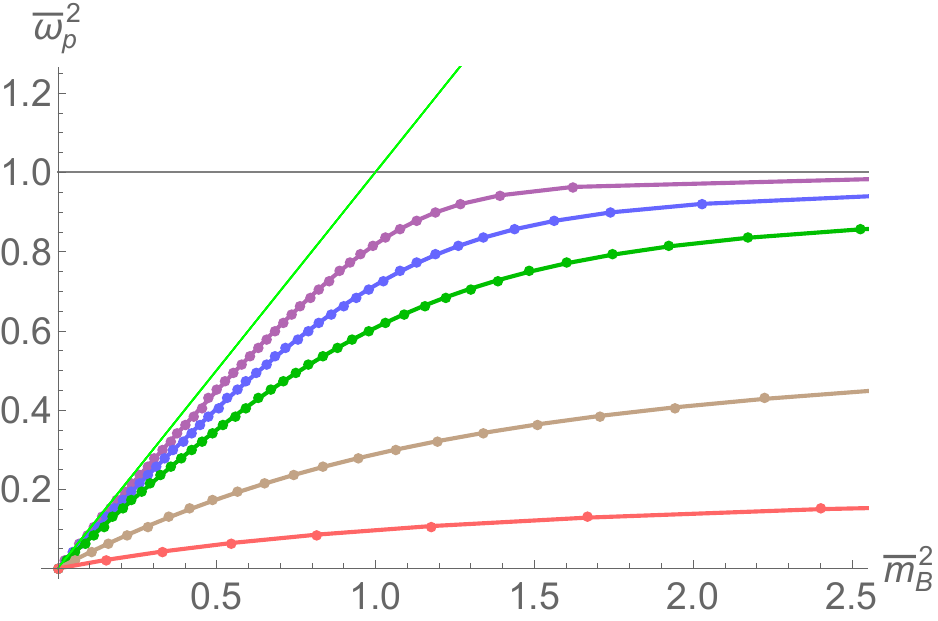}
	\end{center}
\end{minipage}
\caption{Typical bahavior of the polarization tensor at $ q_z =0 $ in the zero-density limit (left). 
The vacuum and total contributions to the polarization tensor are shown 
by gray and purple lines, respectively. 
The plasma frequency as a function of the normalized magnetic-field strength $ \bar m_B^2 $ (right). 
We take $\bar \mu=0  $ and $ \bar T = \{ 0.5, \ 1, \ 3, \ 5, \ 10\} $ from bottom to top. 
}
\label{fig:PF-temp}
\end{figure}

However, for a finite value of $ \bar T  $, the overall coefficient of the divergent term 
deviates from that in the vacuum contribution. 
Even an infinitesimal difference eventually leads to a divergence 
when $ \bar \omega_p^2 $ approaches the threshold closely enough. 
It is important to note that the coefficient of the divergent term in the medium contribution 
is smaller than that of the vacuum contribution, 
because the integrand at finite $  \bar T$ is suppressed by the fermion distribution functions 
as compared to that in the infinite-temperature limit in Eq.~(\ref{eq:integral-infinite-T}). 
Therefore, when $ \bar \omega_p^2  \to 1$ from below, 
the right-hand side of Eq.~(\ref{eq:plasma-frequency}) still goes to negative infinity, i.e., 
\begin{eqnarray}
\lim_{\bar \omega_p \to 1} \big[ - I(\bar \omega_p^2) + \tilde \Pi_ \para^\temp (\bar \omega_p, 0;0) \big] 
= - \infty
\, .
 \end{eqnarray} 
On the other hand, away from the threshold $ \bar \omega_p^2 =1 $, 
the cancellation gets almost complete as we increase $ \bar T $, 
leaving only a small modification to the Schwinger-mass term, i.e., 
\begin{eqnarray}
1 + \big[  - I(\bar \omega_p^2) + \tilde \Pi_ \para^\temp (\bar \omega_p, 0;0) \big]  \sim 1
\, .
 \end{eqnarray} 
The vacuum contribution is comparable in magnitude to the medium contribution, 
and one should include it to get the plasma frequency correctly. 
Those behaviors are shown by the purple line in the left panel of Fig.~\ref{fig:PF-temp}. 
Clearly, one finds an intersection with the linear function for any (positive) value of the slope 
accordingly to the limiting behaviors at $ \bar \omega_p = 0 $ and $ 1 $.

In the right panel of Fig.~\ref{fig:PF-temp}, we show the plasma frequency 
as a function of the normalized magnetic-field strength $  \bar m_B^2$.\footnote{
In practice, one can obtain this plot without numerically searching for the root. 
One only needs to evaluate the integral 
thanks to the simple linear dependence on the left-hand side of Eq.~(\ref{eq:plasma-frequency}). 
} 
We take a set of temperature $ \bar T = \{ 0.5, 1, 3, 5, 10\} $ 
at vanishing chemical potential $ \bar \mu = 0 $. 
All the curves increase as we increase the magnetic-field strength $  \bar m_B^2$. 
The plasma frequency approaches the Schwinger mass, shown with the light green line, 
and saturates at $ \bar \omega_p^2=1 $ as we increase temperature. 
The former behavior is naturally expected from that in the massless limit $ \bar T \to \infty$. 
The latter behavior may be more subtle, implying that an infinitesimal mass makes 
the curve approach $ \bar \omega_p^2=1 $ rather than the Schwinger mass. 
The deviation from the Schwinger mass gets larger as we increase $  \bar m_B^2$, 
and the plasma frequency is strongly suppressed in such a strong-field limit, 
meaning that the fermion-mass effect becomes sizable in the photon dispersion relation.

Here is one caveat; For the LLL approximation to work, 
all the physical scales should be smaller than the magnetic-field strength, 
i.e., $ m^2, \ \omega_p^2, \ T^2 , \ \ll q_fB$. 
Otherwise, the higher Landau levels come into play a role. 
Noticing that $ \bar m_B^2 = \a_\EM \rho_B/m^2 \sim  |q_fB|/m^2 \times 10^{-3} $, 
the LLL approximation should work for the lines shown in Fig.~\ref{fig:PF-temp}, 
except for the regime where $  \bar m_B^2 $ is too small.

\subsubsection{Finite density}

In the finite-density and zero-temperature limit, 
we have $ n_+( m  \bar \epsilon_p  ) =  \theta( \pm \mu - m  \bar \epsilon_p) $. 
In this limit, the integral can be performed as 
\begin{subequations}
\begin{eqnarray}
 \frac12  \int_{-\infty}^\infty \frac{d \bar p_z}{  \bar \epsilon_p }
\frac{  n_+( m  \bar \epsilon_p  )  + n_-( m  \bar \epsilon_p) } {   \bar \epsilon_p^2 - \bar \omega_p^2  }
\= \frac12 \int_{- \sqrt{\bar \mu^2-1}}^{\sqrt{\bar \mu^2-1}} \frac{d \bar p_z}{  \bar \epsilon_p }
\frac{  1 } {  \bar \epsilon_p^2 - \bar \omega_p^2  }
\nnb
\= \frac{1}{ \sqrt{  \bar \omega_p^2  (1- \bar \omega_p^2 ) } } 
\arctan \frac{  \sqrt{\bar \mu^2-1} \, \bar \omega_p^2 }
{   \bar \mu  \sqrt{ \bar \omega_p^2  (1- \bar \omega_p^2 ) } }
\, ,
\label{eq:int-zero-T}
\\
&\to& I(\bar \omega_p^2) \quad {\rm as} \ \ \bar \mu \to \infty
\, ,
\end{eqnarray}
\end{subequations}
Similar to the inifinite-temperature case, the integral reduces to the $ I $ function 
in the infinite chemical potential limit as well, reproducing the massless limit. 
However, the divergent behavior at a finite chemical potential is qualitatively different from 
that in the finite-temperature case. 
This is due to the Pauli-blocking effect in presence of the sharp Fermi surface at zero temperature 
as we have already discussed (see Sec.~\ref{sec:real_part}). 
In Eq.~(\ref{eq:int-zero-T}), the coefficient of the divergent term at $ \bar \omega_p^2 = 1  $ 
does {\it not} depend on $  \bar \mu$, since the arctangent takes $ \pi/2 $ at $ \bar \omega_p^2 = 1  $ 
for any value of $ \bar \mu $ greater than one. 
Therefore, this divergence completely cancels out that from the vacuum contribution 
irrespective of the value of $  \bar \mu$ in Eq.~(\ref{eq:plasma-frequency}), 
and the curve is not closed at $ \bar \omega_p^2 = 1  $ 
in contrast to the finite-temperature case (cf. purple curve in Fig.\ref{fig:PF-temp}). 
The divergence revives once the Fermi surface 
is smeared by effects of an infinitesimal temperature, though.

Because of the complete cancellation of the divergence at zero temperature, 
we need to understand the behavior of the polarization tensor 
beyond the mass threshold $ \bar \omega_p^2 >1 $ as well. 
In this region, the argument of the arctangent in Eq.~(\ref{eq:int-zero-T}) becomes a pure imaginary number. 
It is then useful to use a relation, $  \arctan x = \frac12 i [ \ln (1-ix) - \ln (1+ix) ]$, 
that converts the argument to a real number as 
\begin{eqnarray}
\arctan \frac{ \sqrt{\bar \mu^2-1} \bar \omega_p^2 }
{  \bar \mu  \sqrt{  \bar \omega_p^2  (1- \bar \omega_p^2 ) } }
= 
\ln \frac{  \bar \mu \sqrt{  \bar \omega_p^2  ( \bar \omega_p^2 -1 )  } - \sqrt{\bar \mu^2-1} \, \bar \omega_p^2 }
{  \bar \mu \sqrt{ \bar \omega_p^2  ( \bar \omega_p^2 -1 )  } + \sqrt{\bar \mu^2-1} \, \bar \omega_p^2 }
\, .
\end{eqnarray}
The numerator in the logarithm could take both positive and negative values, 
so that one should compare the magnitudes of those two terms: 
$ \bar \mu^2 \bar \omega_p^2  ( \bar \omega_p^2 -1 )   - ( \sqrt{\bar \mu^2-1} \, \bar \omega_p^2)^2
=  \bar \omega_p^2  ( \bar \omega_p^2  -  \bar \mu^2  ) $. 
The sign changes at $  \bar \omega_p^2 =  \bar \mu^2  $, implying that 
the logarithm diverges at $  \bar \omega_p^2 =  \bar \mu^2  $ 
and has an imaginary part when $ 1 \leq \bar \omega_p^2 \leq \bar \mu^2 $. 
One should add this imaginary part to that from the vacuum contribution $  I ( \bar \omega_p^2) $ 
to find the total expression 
\begin{eqnarray}
\frac{ \bar \omega_p^2 }{ \bar m_B^2 }
\=  1-  \Re e[ I ( \bar \omega_p^2)]
+ \frac{1}{ 2 \sqrt{  \bar \omega_p^2  (\bar \omega_p^2-1 ) } } 
\ln \frac{  \left| \bar \mu \sqrt{  \bar \omega_p^2  ( \bar \omega_p^2 -1 )  } 
- \sqrt{\bar \mu^2-1} \, \bar \omega_p^2 \right| }
{  \bar \mu \sqrt{ \bar \omega_p^2  ( \bar \omega_p^2 -1 )  } + \sqrt{\bar \mu^2-1} \, \bar \omega_p^2 }
\\
&&
- \frac{i \pi }{ 2 \sqrt{  \bar \omega_p^2  (\bar \omega_p^2-1 ) } } 
\ \theta ( \bar \omega_p^2  -  \bar \mu^2 ) 
\nn
\end{eqnarray}
The imaginary parts from the vacuum and medium contributions cancel each other 
in the region $(2m)^2 \leq \omega_p^2 \leq ( 2  \mu )^2  $, 
and the total imaginary part is nonzero only when $ \omega_p^2 \geq ( 2  \mu )^2  $. 
The new threshold is given by the Fermi energy that is nothing but 
the minimum energy required for the pair creation to occur at finite density and zero temperature. 
When $ q_z=0 $, even though the positive-energy antiparticle states are vacant (when $ \mu>0 $), 
twice the Fermi energy is required for the momentum conservation [cf. Eq.(\ref{eq:pair-energy})].

\cout{ 

Because of the absence of the divergence at zero temperature, 
we need to understand the behavior of the polarization tensor in the region $ \bar \omega_p^2 >1 $ as well. 
In this region, the argument of the arctangent in Eq.~(\ref{eq:int-zero-T}) becomes a pure imaginary number. 
It is then useful to use a relation, $  \arctan x = \frac12 i [ \ln (1-ix) - \ln (1+ix) ]$, 
that convert the argument to a real number as 
\begin{eqnarray}
\arctan \frac{ \bar \mu \bar \omega_p^2 }{  \sqrt{  (\bar \mu^2 +1)  \bar \omega_p^2  (1- \bar \omega_p^2 ) } }
= 
\ln \frac{ \sqrt{  (\bar \mu^2 +1)  \bar \omega_p^2  ( \bar \omega_p^2 -1 )  } - \bar \mu \bar \omega_p^2 }
{ \sqrt{  (\bar \mu^2 +1)  \bar \omega_p^2  ( \bar \omega_p^2 -1 )  } + \bar \mu \bar \omega_p^2 }
\, .
\end{eqnarray}
The numerator in the logarithm could take both positive and negative values, 
so that one should compare the magnitudes of those two terms: 
$ (\bar \mu^2 +1)  \bar \omega_p^2  ( \bar \omega_p^2 -1 )   - ( \bar \mu \bar \omega_p^2)^2
=  \bar \omega_p^2  ( \bar \omega_p^2 -1  -  \bar \mu^2  ) $. 
The sign changes at $  \bar \omega_p^2 = 1  +  \bar \mu^2  $, implying that 
the logarithm diverges at $  \bar \omega_p^2 = 1  +  \bar \mu^2  $ 
and has an imaginary part when $ 1 \leq \bar \omega_p^2 \leq 1 +\bar \mu^2 $. 
One should add this imaginary part to that from the vacuum contribution $  I ( \bar \omega_p^2) $ 
to find the total expression 
\begin{eqnarray}
\frac{ \bar \omega_p^2 }{ \bar m_B^2 }
\=  1-  \Re e[ I ( \bar \omega_p^2)]
+ \frac{1}{ 2 \sqrt{  \bar \omega_p^2  (\bar \omega_p^2-1 ) } } 
\ln \frac{ \left|\sqrt{  (\bar \mu^2 +1)  \bar \omega_p^2  ( \bar \omega_p^2 -1 )  } - \bar \mu \bar \omega_p^2\right| }
{ \sqrt{  (\bar \mu^2 +1)  \bar \omega_p^2  ( \bar \omega_p^2 -1 )  } + \bar \mu \bar \omega_p^2 }
\\
&&
- \frac{i \pi }{ 2 \sqrt{  \bar \omega_p^2  (\bar \omega_p^2-1 ) } } 
\ \theta ( \bar \omega_p^2 - 1 -  \bar \mu^2 ) 
\nn
\end{eqnarray}
The imaginary parts from the vacuum and medium contributions cancel each other 
in the region $(2m)^2 \leq \omega_p^2 \leq ( 2 \sqrt{ m^2 + \mu^2} )^2  $, 
and the total imaginary part is nonzero only when $ \omega_p^2 \geq ( 2 \sqrt{ m^2 + \mu^2} )^2  $. 
The lower boundary is given by the Fermi energy that is nothing but 
the minimum energy required for the pair creation to occur at finite density and zero temperature. 
When $ q_z=0 $, even though the positive-energy antiparticle states are vacant (when $ \mu>0 $), 
twice the Fermi energy is required for the momentum conservation [cf. Eq.(\ref{eq:pair-energy})]. 

} 

\begin{figure}
\begin{minipage}{0.45\hsize} 
	\begin{center} 
\includegraphics[width=\hsize]{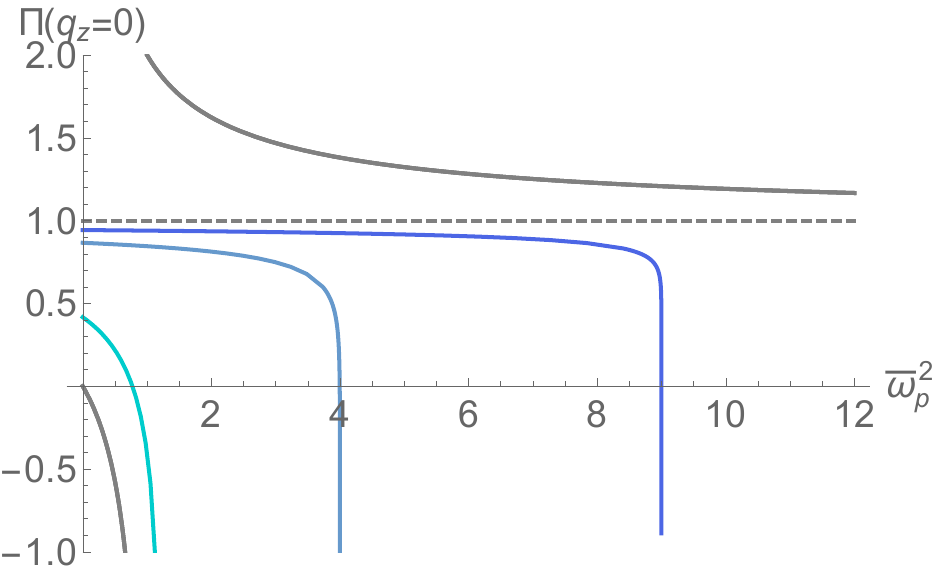}
	\end{center}
\end{minipage}
\begin{minipage}{0.45\hsize}
	\begin{center}
\includegraphics[width=\hsize]{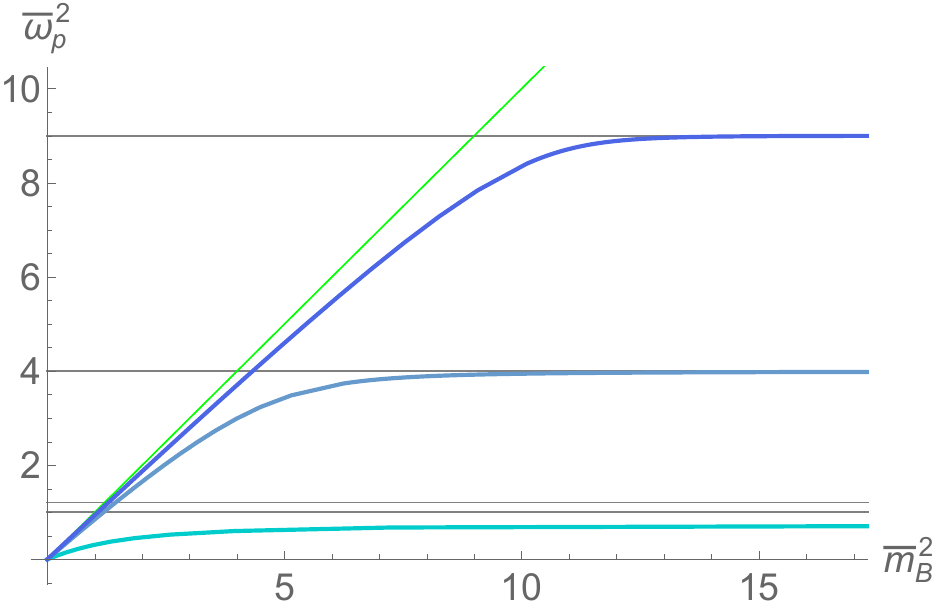}
	\end{center}
\end{minipage}
\caption{The threshold shift in the zero-temperature limit at $ q_z =0 $ (left). 
The vacuum contribution to the polarization tensor is shown by a gray line. 
The plasma frequency as a function of the normalized magnetic-field strength $ \bar m_B^2 $ (right). 
We take $ \bar \mu = \{ 1.1, \ 2 , \ 3 \} $ from bottom to top. 
Note the color correspondences in the plots. 
}
\label{fig:PF-mu}
\end{figure}

We have confirmed the threshold position that depends on $ \bar \mu $ in the zero temperature limit. 
In the left panel of Fig.~\ref{fig:PF-mu}, we show the polarization tensor 
at $ \bar \mu = \{ 1.1 , \ 2 , \ 3 \} $. 
The gray line shows the vacuum contribution as in Fig.~\ref{fig:PF-temp}. 
Those lines at finite $ \bar \mu $ have intersections with the linear function 
for any value of $ \bar m_B^2 $ and $  \bar \mu \ (>1) $. 
In the right panel, we show the plasma frequency as a function of $ \bar m_B^2 $. 
The colors correspond to those in the left panel. 
The plasma frequency approaches the Schwinger mass as we increase $  \bar \mu $. 
As we increase $ \bar m_B^2 $, the plasma frequency saturates 
at $\bar \omega_p^2 =  \bar \mu^2  $ [$ \omega_p^2 =  (2 \mu)^2  $] at zero temperature, 
whereas we have seen that the saturation value is always given by 
$\bar \omega_p^2 =  1 $ [$ \omega_p^2 =  (2 m)^2  $] at finite temperature. 
This difference originates from the shift of the pair-creation threshold induced by the Pauli-blocking effect.

The same caveat mentioned above Eq.~(\ref{eq:int-zero-T}) is applied here. 
The hierarchy, $ m^2, \ \omega_p^2, \ \mu^2 , \ \ll q_fB$, 
should be satisfied for the LLL approximation to work. 
The LLL approximation should work for the lines shown in Fig.~\ref{fig:PF-mu}, 
except for the regime where $  \bar m_B^2 $ is too small.

\cout{
\begin{subequations}
\begin{eqnarray}
y \= \frac{ \bar \omega_p^2 }{ \bar m_B^2 }
\, ,
\\
y \=  \cgd{-}  \bigg[ \, 1- I ( \bar \omega_p^2)
+ \frac12  \int_{-\infty}^\infty \frac{d p_z}{  \bar \epsilon_p }
\frac{  n_+( m  \bar \epsilon_p  )  + n_-( m  \bar \epsilon_p) } {   \bar \epsilon_p^2 - \bar \omega_p^2  }
 \,  \bigg] 
 \, .
\end{eqnarray}
\end{subequations}

Once we generate a set of $ (\bar \omega_p^2 , y) $ by evaluating the integral 
for given values of $ \bar T $ and $ \bar \mu $, 
the magnetic-field dependence of the plasma frequency is given 
as $ (\bar m_B^2, \bar \omega_p^2) =  ( \bar \omega_p^2/y ,  \bar \omega_p^2)$. 
Likewise, for a given value of $ \bar m_B^2 $, one can make plots of the plasma frequency versus 
temperature and chemical potential as 
$ ( \bar T, \bar \omega_p^2) =  (  \bar T ,  \bar m_B^2 y)$ 
and $ ( \bar \mu, \bar \omega_p^2) =  (  \bar \mu ,  \bar m_B^2 y)$, respectively. 
}

\subsection{Comments on a global structure of the photon dispersion relation}

In the previous subsection, we investigated the plasma frequency that is obtained 
as the solution to Eq.~(\ref{eq:plasma-frequency}). 
We have searched for the solutions within the low-frequency regime 
below the pair-creation threshold where the polarization tensor is a real-valued function. 
However, as we inspected in Sec.~\ref{sec:medium}, 
the polarization tensor takes a finite value above the threshold 
and moreover becomes a complex-valued function.

These facts suggest existence of another complex solution for the plasma frequency. 
Its real part may be larger than the threshold energy, 
while its imaginary part implies a finite lifetime of the excitation. 
The presence of such an unstable branch was indeed shown in Ref.~\cite{Hattori:2012ny} 
in the absence of medium effects. 
It would be interesting to investigate the medium effects on the unstable branch. 
It is also important to extend the solution search to a finite photon momentum $ q_z $ 
in order to uncover the entire global structure of the photon dispersion relation. 
We leave those projects as the future works.

\section{Summary and discussions}

In the first paper of the series, we provided detailed account of 
computing the in-medium polarization tensor in the strong magnetic fields. 
We also showed the resummed photon propagator with the polarization tensor. 
The polarization-dependent pole positions explicitly indicate 
the occurrence of the magneto-birefringence. 

We then discussed the physical origins of the peak structures 
in the real and imaginary parts of the polarization tensor. 
They basically originate from the pair creation from a single photon 
and the Landau damping. 
There is a delicate interplay between the vacuum and medium contributions 
to the pair creation due to the Pauli-blocking effect and the reciprocal process, 
i.e., the pair annihilation of thermal particles. 
These points are well understood by looking at the finite-density and zero-temperature case. 
We provided a complete analytic expression for this case and discussed its essential behaviors; 
The threshold positions are shifted from the fermion mass scale to the chemical potential 
due to the Pauli-blocking effect inside the sharp Fermi surface. 
Finite-temperature effects smear the sharp behaviors.

Such kinematics and the medium effects are best captured by the imaginary part 
of the polarization tensor that provides the cross-section of the physical processes 
according to the optical theorem. 
Moreover, we have provided the analytic expression for the imaginary part 
for arbitrary temperature and density as long as they are smaller than the magnetic-field strength. 
Therefore, one may apply it to the optical activities including the photon emission/absorption rates 
in the quark-gluon plasma (see also 
Refs.~\cite{Wang:2020dsr, Wang:2021ebh, Wang:2021eud} for recent works) 
and in neutron stars/magnetars \cite{Mignani:2016fwz, Enoto:2019vcg}. 
It should be noticed that such processes generate photon polarizations
that can be an evidence for the presence of strong magnetic fields.

We finally investigated the photon masses called the plasma frequency and the Debye mass. 
They have different definitions due to the noncommutative limits of 
the vanishing photon energy and momentum. 
The Debye mass is an important quantity that determines the in-medium interaction strength. 
Such information is necessary for computing, e.g., the transport coefficients in magnetic fields 
\cite{Hattori:2016lqx, Hattori:2016cnt, Hattori:2017qih, Fukushima:2017lvb, Fukushima:2019ugr}. 
On the other hand, the plasma frequency is the mass gap in 
the in-medium photon dispersion relation. 
It captures the magneto-birefringence that is an outcome of 
the interplay between the vacuum birefringence and the medium corrections. 
The full global structure of the photon dispersion relation was investigated 
in the vacuum case \cite{Hattori:2012ny}, 
but is yet to be investigated with the medium contribution elsewhere.

\vspace{1cm}

{\it Acknowledgments}.--- 
This work is supported in part by JSPS KAKENHI under grant Nos.~20K03948 and 22H02316.

\appendix

\section{Computation of the polarization tensors}

\subsection{Vacuum contribution}

\label{sec:comp-vacuum}

We briefly recapitulate the computation of the vacuum contribution. 
Inserting the explicit form of the LLL propagator that is 
factorized into the parallel and perpendicular parts, 
we find that the one-loop vacuum polarization tensor is also factorized as 
\begin{eqnarray}
\label{eq:VV}
\Pi^{\mu\nu}_{R} (q) =   I_\perp (q_\perp^2)  \Pi_{1+1}^{\mu\nu} (q_\parallel)
\, ,
\end{eqnarray}
where
\begin{subequations}
\begin{eqnarray}
I_\perp (q_\perp^2) &=& 2^2 \int \!\! \frac{d^2p_\perp}{(2\pi)^2} 
e^{ - \frac{ 1 }{ |q_f B| } (|\bp_\perp|^2 + |\bp_\perp + \bq_\perp |^2 ) } 
\, ,
\\
i \Pi_{1+1}^{\mu\nu} (q_\parallel) &=& - (-iq_f)^2 \int \!\! \frac{d^2p_\parallel}{(2\pi)^2} 
\frac{\tr [\,  \gam^\mu_\parallel i (\sla p_\parallel + m) \prj_+
\gam_\parallel^\nu i ( \, (\sla p_\parallel + \sla q_\parallel) + m\, ) \prj_+ \, ]}
{ (p_\parallel^2 -m^2)( \,  (p_\parallel + q_\parallel)^2 -m^2 \,)}
\, .
\end{eqnarray}
\end{subequations}
Here, the spinor trace is taken at the four dimensions. 
Performing the elementary integral for the transverse momentum, 
we find 
\begin{eqnarray}
I_\perp (q_\perp^2) = \rho_B  e^{ - \frac{ |\bq_\perp|^2 }{2 |q_f B| } }
\, ,
\label{eq:I_Gaussian}
\end{eqnarray}
where the Landau degeneracy factor $ \rho_B =  |q_f B | /( 2\pi) $ appeared from the integral.

On the other hand, the longitudinal part is the polarization tensor in the (1+1)-dimensional QED, 
i.e., the Schwinger model \cite{Schwinger:1962tp}. 
After the standard treatment by the use of the Feynman parameter, we find 
\begin{eqnarray}
\label{eq:int_para0}
i \Pi_{1+1}^{\mu\nu} (q_\parallel) &=&  - q_f^2 \frac{4}{2}  \int_0^1 \!\! dx \left[ \, 
\int \!\! \frac{d^2\ell_\parallel}{(2\pi)^2} 
\frac{ 2\ell_\parallel^\mu \ell_\parallel^\nu - \ell_\parallel^2 g_\para^{\mu\nu}}{(\ell_\parallel^2 - \Delta_\para)^2}
+ 
\int \!\! \frac{d^2\ell_\parallel}{(2\pi)^2} 
\frac{ \alpha^{\mu\nu} }{(\ell_\parallel^2 - \Delta_\para)^2}
\, \right]
\, , 
\end{eqnarray}
where $ \ell_\para^\mu = p_\para^\mu + x(1-x) q_\para^\mu $, 
$\Delta _\para = m^2 - x(1-x)q_\para^2  $, and $ \alpha^{\mu\nu} =
m^2 g_\parallel^{\mu\nu} + x(1-x) ( q_\parallel^2 g_\parallel^{\mu\nu} - 2 q_\parallel^\mu q_\parallel^\nu)$. 
One needs to be careful about the gauge-invariant regularization. 
In the dimensional regularization, the numerator of the first integral is proportional to 
$ \epsilon_2 g_\parallel^{\mu\nu}  $ with $  \epsilon_2 \equiv  (2-d)$, 
which appears to vanish in the two dimensions. 
However, a factor of $  1/\epsilon_2$ arises from the divergent integral. 
The leading term provides a finite and gauge-invariant result: 
\begin{eqnarray}
\Pi_{1+1}^{\mu\nu} (q_\parallel) &=&  - \frac{ q_f^2}{\pi}  
( q_\parallel^2 g_\parallel^{\mu\nu} -  q_\parallel^\mu q_\parallel^\nu)
\int_0^1 \!\! dx x(1-x) \Delta_\para^{-1} 
\label{eq:int_para}
\, .
\end{eqnarray}
Performing the elementary integral, we obtain the result in Eq.~(\ref{eq:vac_massive}). 
 

\subsection{Medium contribution}

\label{sec:comp-medium}

Next, we show the computation of the thermal contribution at finite temperature and density in detail. 
Inserting the real-time propagators (\ref{eq:S-all}) into Eq.~(\ref{eq:med1}), we have 
\begin{eqnarray}
i \M_1^{\mu\nu}
&=& -   q_f ^2 \frac{ |q_f B| }{8\pi} e^{- \frac{ \vert \bq_\perp \vert ^2}{2 |q_f B| } }
i \int \frac{d p_z}{2\pi} \frac{ 1 }{ \epsilon_p}
\biggl[\, 
n_+(\epsilon_p) \frac{\tr[ \gam^\mu_\parallel  ( \slashed p_+ + \slashed q_\parallel )
\gam^\nu_\parallel   \slashed p _{+}   ] + 4m^2 g^{\mu\nu}_\parallel }
{ (p_++q_\parallel)^2 - m^2 -i \sgn(\epsilon_p+q^0) \varepsilon }
\nonumber
\\
&&\hspace{4.3cm} + n_-(\epsilon_p)
\frac{\tr[ \gam^\mu_\parallel   ( \slashed p_- + \slashed q_\parallel)
\gam^\nu_\parallel   \slashed p _{-}  ]
+ 4m^2 g^{\mu\nu}_\parallel  }
{ (p_-+q_\parallel)^2 - m^2 -i \sgn(-\epsilon_p+q^0) \varepsilon }
\, \biggl]
\, ,
\end{eqnarray}
where $p_{ \pm } = ( \pm \epsilon_p , 0, 0, p_z)$. 
We just consumed the delta functions with the $ p^0 $ integral 
and used a relation $ n_\pm (-x) = 1- n_\mp (x)   $ to drop the vacuum part. 
In the same manner, one can obtain the other contribution 
\begin{eqnarray}
i\M_2^{\mu\nu}
&=& -   q_f ^2 \frac{ |q_f B| }{8\pi} e^{- \frac{ \vert \bq_\perp \vert ^2}{2  |q_f B|  } }
i \int \frac{d p_z}{2\pi} \frac{ 1 }{ \epsilon_p}
\biggl[\, 
n_+(\epsilon_p) \frac{\tr[ \gam^\mu_\parallel \slashed p _{+} 
\gam^\nu _\parallel ( \slashed p_+ - \slashed q_\parallel) ] 
+ 4m^2 g^{\mu\nu}_\parallel }
{ (p_+-q_\parallel)^2 - m^2  +  i \sgn(\epsilon_p-q^0) \varepsilon }
\nonumber
\\
&&\hspace{4.3cm} 
+ n_-(\epsilon_p) \frac{\tr[ \gam^\mu_\parallel \slashed p _{-} 
\gam^\nu_\parallel ( \slashed p_- - \slashed q_\parallel ) ] 
+ 4m^2 g^{\mu\nu}_\parallel }
{ (p_- - q_\parallel)^2 - m^2  +  i \sgn(-\epsilon_p-q^0) \varepsilon }
\, \biggl]
\, ,
\end{eqnarray}
where we shifted the integral variables as 
$ p_\para^\mu \to p_\para^{\prime \mu} =  p_\para^\mu + q_\para^\mu $. 
This shift is allowed since the integral is finite. 
Notice that changing the integral variable $p_z \to - p_z$ 
induces a change $ p_{ \pm} \to (\pm \epsilon_p, 0, 0, -p_z) =  - p_{ \mp} $ 
and also that the spinor trace is symmetric, i.e., $\tr[\gam^\mu \slashed p_1 \gam^\nu \slashed p_2]
= \tr[\gam^\mu \slashed p_2 \gam^\nu \slashed p_1]$ for arbitrary vectors $ p_{1,2}^\mu $. 
It follows from those observations that $  \M_2^{\mu\nu}  $ has the same form as $   \M_1^{\mu\nu} $ 
up to simultaneous replacements $ n_\pm (\epsilon_p) \to n_\mp (\epsilon_p) $. 
Therefore, the sum of the two contributions reads 
\begin{eqnarray}
\label{eq:split-M}
\Pi_\temp^{\mu\nu} =    \M_1^{\mu\nu} +  \M_2^{\mu\nu} 
=  -  q_f ^2 \frac{ |q_f B| }{2\pi} e^{- \frac{ \vert q_\perp \vert ^2}{2 |q_f B| } } \M_\para^{\mu\nu}
\, ,
\end{eqnarray}
where the fermion distribution functions are factorized in the additive form as 
\begin{eqnarray}
\M^{\mu\nu}_\para \= 
\frac14 \int \frac{d p_z}{2\pi} \frac{ n_+(\epsilon_p) +n_-(\epsilon_p)  }{ \epsilon_p}
\nnb 
&& \times
 \biggl[\, 
\frac{\tr[ \gam^\mu_\parallel ( \slashed p_+ - \slashed q_\parallel )
\gam^\nu_\parallel  \slashed p _{ + }  ] 
+ 4m^2 g^{\mu\nu}_\parallel }{ (p_+ - q_\parallel)^2 - m^2   }
+ 
\frac{\tr[ \gam^\mu_\parallel ( \slashed p_- - \slashed q_\parallel )
\gam^\nu _\parallel   \slashed p _{-}  ] 
+ 4m^2 g^{\mu\nu}_\parallel } { (p_- -q_\parallel)^2 - m^2    }
\, \biggl]
\label{eq:M_para}
\, .
\end{eqnarray}
The explicit forms of the denominators are given as 
$ ( p_\pm - q_\parallel)^2 -m ^2 =  2 ( \mp \omega \epsilon_p + q_z p_z + \frac{1}{2} q_\parallel^2 ) $ 
with $q_\parallel = (\omega, 0, 0, q_z)$. 
Putting the whole integrand over a common denominator, 
the denominator reads 
\begin{eqnarray}
\{( p_+ - q)_\parallel^2 -m ^2\} \{( p_- - q)_\parallel^2 -m ^2\} 
&=& - 4 \{ q_\parallel^2 ( p_z - \frac{1}{2} q_z) ^2
- \frac{1}{4} \omega^2 (q_\parallel^2 -4m^2) \}
\, .
\end{eqnarray}

Next, we perform the spinor trace in the numerators. 
We examine the diagonal and off-diagonal Lorentz components separately. 
In the diagonal components, the trace is carried out as 
\begin{eqnarray}
T_\pm ^\diag =
\frac14 \tr[ \gam^\mu_\parallel ( \slashed p_\pm - \slashed q_\parallel  )
\gam^\nu_\parallel  \slashed p _{ \pm}  ]
= ( \epsilon_p^2 + p_z^2  - q_z p_z )  \mp \omega \epsilon_p 
\, .
\end{eqnarray}
After the reduction to common denominator, the numerator is arranged as 
\begin{eqnarray}
&& \hspace{-1cm}
T_+^\diag  \{( p_- - q)_\parallel^2 -m ^2\} + T_-^\diag  \{( p_+ - q)_\parallel^2 -m ^2\} 
\nnb
\= 8 q_z p_z \{ (p_z- \frac{1}{2} q_z )^2 - \frac{1}{4} \omega^2 \}
+ 4m^2 \{ q_z p_z - \omega^2 + \frac{1}{2} q_\parallel^2 \} 
\, .
\end{eqnarray}
As we will see shortly, the mass-independent part vanishes identically 
when integrated with respect to $p_z$. 
Likewise, the trace for the off-diagonal (0,3) component is carried out as 
\begin{eqnarray}
T_\pm^{03} = \frac14 \tr[ \gam^0_\parallel ( \slashed p_\pm - \slashed q_\parallel  )
\gam^3_\parallel  \slashed p _{ \pm}  ] 
=  - \omega p_z  \pm ( 2 \epsilon_p p_z -  q_z \epsilon_p )
\, ,
\end{eqnarray}
leading to the corresponding numerator 
\begin{eqnarray}
&& \hspace{-1cm}
T_+^{03}  \{( p_- - q)_\parallel^2 -m ^2\}  + T_-^{03}  \{( p_+ - q)_\parallel^2 -m ^2\} 
\nnb
&= &
8 \omega p_z \{ (p_z- \frac{1}{2} q_z )^2 - \frac{1}{4} \omega^2 \}
+ 4 m^2 \{ 2 \omega p_z - \omega q_z \}
\, .
\end{eqnarray}
The trace should be symmetric, i.e., $ T_\pm^{03} = T_\pm^{30} $, 
because there is no room for the completely antisymmetric tensor to appear without $ \gam^5 $.

Now, one can show that the medium correction to 
the polarization tensor vanishes in the massless case. 
Both the numerator and denominator are 
proportional to $ (p_z- \frac{1}{2} q_z )^2 - \frac{1}{4} \omega^2  $ to cancel each other. 
What remains is a simple integral 
\begin{eqnarray}
\left.  \M_\para \right|_{m=0} 
&\propto& \int_{-\infty}^\infty \frac{d p_z}{2\pi} \frac{ n_+(\epsilon_p) + n_-(\epsilon_p) }{ \epsilon_p} p_z = 0
\, .
\end{eqnarray}
Clearly, this integral is vanishing with $\epsilon_p =  |p_z| $ at $ m=0 $.\footnote{
The correct dispersion relation of the massless LLL fermions is 
$\epsilon_p =  \pm p_z $ without the symbol of absolute value. 
Nevertheless, the spectral function $ \rho(p) \propto \left[ \delta( p^0  - p_z) + \delta( p^0 + p_z) \right] /p^0  $ 
is still an even function of $ p_z $, allowing a replacement $  p_z \to |p_z| $ in $ \epsilon_p  $. 
Thus, the conclusion still holds with the correct dispersion relation. 
} 

%

Even in the massive case, one finds a partial cancellation between the numerator and denominator. 
As anticipated from the complete cancellation in the massless case, 
the residual terms are all proportional to the fermion mass $ m^2 $. 
Including the terms proportional to $ 4m^2 g^{\mu\nu}_\parallel  $ in Eq.~(\ref{eq:M_para}) as well, 
we finally arrive at 
\begin{eqnarray}
\label{eq:722}
\M_\para^{\mu\nu} 
=  m^2 P_\para^{\mu\nu} 
\!\! \int_{-\infty}^\infty \frac{d p_z}{2\pi \epsilon_p} 
\frac{ ( q_\parallel^2 + 2 q_z p_z) \,   [n_+(\epsilon_p)  + n_-(\epsilon_p)] } 
{  q_\parallel^2  ( p_z - \frac{1}{2} q_z) ^2 - \frac{\omega^2}{4 }  (q_\parallel^2 -4m^2) } 
\, .
\end{eqnarray}
Notice that the diagonal and off-diagonal components are combined 
in the form of $ P_\para^{\mu\nu}  $, indicating manifest transversality of the polarization tensor, 
serving as a consistency check of the calculation. 
The above leads to the result in Eq.~(\ref{eq:Pi_medium}).

\section{Analytic integration at finite density: Threshold shifts}

\label{sec:zero-T}

In this appendix, we perform the integral for the medium contribution (\ref{eq:Re-zero-T}) 
at finite density and zero temperature. 
The Fermi-Dirac distribution functions reduce to the step functions 
that provide the integral with the cutoff as  
\begin{eqnarray}
\label{eq:Re-zero-T}
\left.  \tilde  \Pi_\para^\temp (\omega, q_z) \right|_{T=0}
= \frac{  m^2    }{ q_\para^2\sqrt{1- 4m^2/q_\para^2 }  }
\int_{-p_F}^{p_F} \frac{d p_z}{ \epsilon_p} 
\Big( \frac{ \epsilon_p^+ s_+ }{ p_z - p_z^+ } - \frac{ \epsilon_p^- s_-}{ p_z - p_z^- } \Big)
\, ,
\end{eqnarray}
where $ p_F = \sqrt{ \mu^2 - m^2} $ is the Fermi momentum at zero temperature. 
Now, the integral can be performed by the use of an indefinite integral 
\begin{eqnarray}
 \int \frac{d p_z}{\epsilon_p} \frac{1}{p_z - a}
 = \frac{1}{ \sqrt{ a^2 + m^2}}
  \arctanh \left( - \frac{ a p_z + m^2 }{ \sqrt{ (p_z^2 + m^2) (a^2 +m^2) } } \right)
  + C
  \, ,
\end{eqnarray}
where $  C$ is the integral constant. 
Applying this formula, we obtain 
\begin{eqnarray}
\label{eq:integrals-zero-density}
 \int_{ -p_F}^{p_F} \frac{d p_z}{\epsilon_p} \frac{1}{p_z - p_z^\pm}
 \= \frac{1}{  \epsilon_p^\pm} \Big[
  \arctanh \left( - \frac{  p_z^\pm p_F + m^2 }{ \mu  \epsilon_p^\pm } \right)
  -
    \arctanh \left( - \frac{  - p_z^\pm p_F+ m^2 }{ \mu \epsilon_p^\pm } \right)
  \Big]
  \nnb
\= - \frac{1}{2 \epsilon_p^\pm} \Big[
\ln \left(  \frac{  \mu \epsilon_p^\pm + ( p_z^\pm p_F + m^2 ) } 
{  \mu \epsilon_p^\pm- ( p_z^\pm p_F  + m^2 ) } \right)
- 
\ln \left( \frac{  \mu \epsilon_p^\pm + ( - p_z^\pm p_F + m^2 ) } 
{  \mu \epsilon_p^\pm - ( - p_z^\pm  p_F  + m^2 ) } \right)
  \Big]
  \nnb
 \=  \frac{1}{ \epsilon_p^\pm} 
\ln \frac{ \mu p_z^\pm -  p_F \epsilon_p^\pm } {  \mu p_z^\pm + p_F  \epsilon_p^\pm  } 
 \, ,
\end{eqnarray}
where we used a relation, $\arctanh( z)   = - \frac12 \{ \ln (1- z) - \ln(1+z) \}  $. 
Plugging this result into Eq.~(\ref{eq:Re-zero-T}), we arrive at a rather simple form 
\begin{eqnarray}
\left. \tilde \Pi_\para^{\temp }(\omega, q_z) \right|_{T=0}
\=  \frac{ m^2  }{ q_\para^2\sqrt{1- 4m^2/q_\para^2 }  } 
\Big[ 
s_+ \ln  \frac{ \mu  p_z^+ - p_F \, \epsilon_p^+ }  
{ \mu p_z^+ + p_F \, \epsilon_p^+  } 
-
s_- \ln  \frac{ \mu p_z^- - p_F \, \epsilon_p^- }  
{ \mu p_z^- + p_F \, \epsilon_p^-  } 
\Big]
\label{eq:Pi_zero-T}
\, .
\end{eqnarray}

We examine some crucial properties of the above result. 
When $ 1 - 4m^2 /q_\para^2 <0 $, that is, $ 0 < q_\para^2 < 4m^2 $, 
we have complex-conjugate properties 
$ (\epsilon_p^\pm)^\ast =  \epsilon_p^\mp $ and $ (p_z^\pm)^\ast = p_z^\mp $, 
and so do the arguments of the logarithms. 
Remember also that $ s_\pm=1 $ when $ 1 - 4m^2 /q_\para^2 <0 $ 
as promised above Eq.~(\ref{eq:original}). 
Therefore, the difference between the logarithms is a pure imaginary value. 
Accordingly, in this region, the medium contribution is a real-valued function 
\begin{eqnarray}
\left. \tilde \Pi_\para^{\temp }(\omega, q_z) \right|_{T=0}
\=  \frac{ 2 m^2  }{ q_\para^2\sqrt{ |1- 4m^2/q_\para^2| }  } 
\arg \Big(   \frac{ \mu  p_z^+ - p_F \, \epsilon_p^+ }  { \mu p_z^+ + p_F \, \epsilon_p^+  }  \Big)
\label{eq:Pi_zero-T-real-app}
\, ,
\end{eqnarray}
where ``$ \arg  $'' denotes the argument of the complex variable. 
We take the principal value within $ [-\pi,\pi] $. 
When the photon momentum approaches the boundaries 
as $ q_\para^2 \to 0 $ from above and $ q_\para^2 \to 4m^2 $ from below, 
$ \tilde \Pi_\para^{\temp } $ approaches zero and positive infinity, respectively. 
The latter divergent behavior is a remarkable property specific to 
the case of nonzero density at zero temperature, 
and exactly cancels the divergence from the $  I$ function in the vacuum contribution. 
These properties can be confirmed as follows. 
The real and imaginary parts are arranged as 
\begin{subequations}
\begin{eqnarray}
\Re e \Big [  \frac{ \mu  p_z^+ - p_F \, \epsilon_p^+ }  { \mu p_z^+ + p_F \, \epsilon_p^+  } \Big]
  \=  \frac{\mu ^2 q_z^2 - p_F^2 \omega^2 +(\mu ^2 \omega^2 - p_F^2 q_z^2 ) 
  ( 4m^2 - q_\para^2)/ q_\para^2  }
{ 4 | \mu p_z^+ + p_F \, \epsilon_p^+|^ 2 }  
\, ,
\\
\Im m  \Big[ \frac{ \mu  p_z^+ - p_F \, \epsilon_p^+ }  { \mu p_z^+ + p_F \, \epsilon_p^+  } \Big]
  \=  \frac{2  \mu p_F}   { 4 | \mu p_z^+ + p_F \, \epsilon_p^+|^ 2 } 
\,   \sqrt{  q_\para^2 (4m^2 - q_\para^2  ) }
  \, .
\end{eqnarray}
\end{subequations}
When $ q_\para^2 \to 0  $ from above, 
the imaginary part approaches zero from above, 
while the real part diverges $ \omega^2 (\mu ^2   - p_F^2)  \frac{4m^2}{q_\para^2} \to + \infty $. 
On the other hand, when $ q_\para^2 \to 4m^2  $ from below, 
the imaginary part approaches zero from above, while the real part approaches a negative value 
\begin{eqnarray}
\mu ^2 q_z^2 - p_F^2 \omega^2 
= - q_\para^2 \mu^2 + m^2 \omega^2
\leq -  \omega^2 (\mu^2 - m^2)
\, .
\end{eqnarray}
The rightmost side is negative as long as $ \mu \geq m $. 
The argument in Eq.~(\ref{eq:Pi_zero-T-real-app}) evolves from 0 to $ +\pi $ as 
we increase the photon momentum from $  q_\para^2 =0 $ to $ 4m^2$. 
Therefore, the medium contribution (\ref{eq:Pi_zero-T-real-app}) diverges 
to positive infinity as $q_\para^2 \to 4m^2  $. 
This divergence exactly cancels that from the vacuum contribution (\ref{eq:threshold-vac}). 
Since the divergent is associated with the threshold behavior of the pair creation, 
the absence of the divergence implies that the threshold is shifted to higher energies.

We look for positions of the shifted pair-creation threshold by directly inspecting the imaginary part 
in the region $ q_\para^2 \geq 4m^2 $, as well as in $ q_\para^2 \leq 0 $ for the Landau damping. 
When $ 1 - 4m^2 /q_\para^2 > 0 $, that is, $ q_\para^2 <0 $ or $ 4m^2 \leq q_\para^2 $, 
$ \epsilon_p^\pm $ and $ p_z^\pm $ take real values. 
In this region, the arguments of the logarithms can take both positive and negative values, 
so that one can find imaginary parts when the arguments become negative. 
Notice also that $ \tilde \Pi_\para^{\temp } $ is an even function of $ q_z $ for the parity invariance, 
which can be confirmed with the transformation properties 
$ p_z^\pm \to - p_z^\mp $, $ \epsilon_p^\pm \to \epsilon_p^\mp $, 
and $ s_\pm \to s_\mp $ when $ q_z \to - q_z $. 
Below, we assume that $ q_z $ is a semi-positive quantity without loosing generality.

The logarithms yield imaginary parts when 
\begin{eqnarray}
 ( \mu p_z^\pm)^2 - ( p_F \epsilon_p^\pm)^2
=   m^2 ( \epsilon_p^\pm + \mu) ( \epsilon_p^\pm - \mu)
< 0 
\, .
\end{eqnarray}
Thus, the thermal contribution to the polarization tensor 
acquires an imaginary part when $ \epsilon_p^\pm < \mu $. 
The signs of the imaginary parts depend on 
the signs of the infinitesimal imaginary parts of the arguments. 
One should keep track of the signs of the linear terms in the infinitesimal displacement 
\begin{eqnarray}
\left. \frac{ \mu  p_z^\pm -  p_F \epsilon_p^\pm } { \mu p_z^\pm + p_F  \epsilon_p^\pm  } 
\right|_{\omega \to \omega+ i \epsilon}
\= 
\left. \frac{ \mu p_z^\pm -  p_F \epsilon_p^\pm } {  \mu p_z^\pm + p_F  \epsilon_p^\pm  } 
\right|_{\epsilon=0}
\pm  i \epsilon s_\pm 
\frac{ \omega  \mp q_z \sqrt{ 1 - \frac{4m^2}{q_\para^2}}   } 
{  q_\para^2 \sqrt{ 1 - \frac{4m^2}{q_\para^2}}  }
+\order(\epsilon^2)
\nnb
\= 
\left. \frac{ \mu  p_z^\pm -  p_F \epsilon_p^\pm } {  \mu p_z^\pm + p_F  \epsilon_p^\pm  } 
\right|_{\epsilon=0}
\pm i \epsilon s_\pm  s_\mp \sgn(  q_\para^2  ) 
+\order(\epsilon^2)
\end{eqnarray}
We have dropped all the positive-definite factors since we are only interested in 
the signs in front of the linear term in the infinitesimal parameter $ \epsilon $.  
Including the sign functions $ \pm s_\pm $ in front of the logarithms in Eq.~(\ref{eq:Pi_zero-T}), 
the signs of the imaginary parts are found to be 
\begin{eqnarray}
\pm s_\pm [ \pm s_\pm s_\mp  \sgn( q_\para^2 ) ]
=   s_\mp  \sgn(  q_\para^2  )
=   s_\pm
\end{eqnarray} 
We used $ s_\pm $ given in Eq.~(\ref{eq:signs}).

According to the above signs, the imaginary part of the medium contribution is obtained as 
\begin{eqnarray}
\Im m [\tilde \Pi_\para^{\temp } ]
\=  \frac{  m^2  }{ q_\para^2\sqrt{1- 4m^2/q_\para^2 }  } 
\times \pi \Big [ \{ \theta(\mu - \epsilon_p^+) +  \theta(\mu - \epsilon_p^-) \}  
\theta(q_\para^2 -4m^2) \sgn(\omega)
\nnb
&&
+ \{ \theta(\mu - \epsilon_p^+)  -  \theta(\mu - \epsilon_p^-) \}  \theta(-q_\para^2)  \Big ]
\label{eq:Pi_medium-zeroT-imag}
\, .
\end{eqnarray} 
Including the vacuum contribution, the total imaginary part reads 
\begin{eqnarray}
\Im m [\tilde \Pi_\para ]
\=  \frac{   - 2 \pi m^2  }{ q_\para^2\sqrt{1- 4m^2/q_\para^2 }  } 
\Big[\, \big[1  - \frac12 \{\theta(\mu - \epsilon_p^+) +  \theta(\mu - \epsilon_p^-)\} \big] 
\theta(q_\para^2 -4m^2)   \sgn(\omega) 
\nnb
&& 
 - \frac12 \{ \theta(\mu - \epsilon_p^+)  -  \theta(\mu - \epsilon_p^-)\}  \theta(-q_\para^2) \, \Big]
\label{eq:Pi_total-zeroT-imag}
\, .
\end{eqnarray} 
The kinematics is made clear if one arranges the step functions as 
\begin{subequations}
\begin{eqnarray}
\label{eq:imag-density-pair}
1  - \frac12 \{\theta(\mu - \epsilon_p^+) +  \theta(\mu - \epsilon_p^-)\} 
\= \frac12 \{\theta( \epsilon_p^+ - \mu) +  \theta( \epsilon_p^- - \mu) \} 
\, ,
\\
- \frac12 \{\theta(\mu - \epsilon_p^+) -  \theta(\mu - \epsilon_p^-)\}
\= \frac12 [\,  \theta(\mu - \epsilon_p^-)  \theta( \epsilon_p^+ - \mu )
-  \theta(\mu - \epsilon_p^+)  \theta( \epsilon_p^- - \mu )   \, ]
 \, .
\end{eqnarray}
\end{subequations}
Those expressions agree with the previous results in Eqs.~(\ref{eq:N+_T=0}) and (\ref{eq:N-_T=0}). 
The imaginary part in the vacuum contribution is completely cancelled 
by that from the medium contribution in the region between 
the original threshold at $q_\para^2 =  4m^2 $ and the new threshold where 
the first kinematical condition $\mu < \epsilon_p^-  $ is satisfied. 
As is clear in Eq.~(\ref{eq:imag-density-pair}), a half of the vacuum contribution is restored 
until the other condition $\mu < \epsilon_p^+  $ is met. 
When $\mu < \epsilon_p^+  $, the vacuum contribution is fully restored 
since there is no Pauli-blocking effect above the Fermi energy at zero temperature. 
The pair-creation channels open sequentially as we increase $ q_\para^2 $. 
When $  q_z =0 $, the two thresholds degenerate into each other.

In the infinite-density limit ($ \mu/m \to \infty $), one can confirm that 
the peaks associated with the pair-creation threshold and the Landau damping go away. 
In this limit, we have 
\begin{eqnarray}
\lim_{\mu/m \to \infty} \left. \tilde \Pi_\para^{\temp }(\omega, q_z) \right|_{T=0}
\=  \frac{ m^2  }{ q_\para^2\sqrt{1- 4m^2/q_\para^2 }  } 
\Big[ 
s_+ \ln  \frac{ p_z^+ -  \epsilon_p^+ }  { p_z^+ +  \epsilon_p^+  } 
-
s_- \ln  \frac{ p_z^- -  \epsilon_p^- }  { p_z^- +  \epsilon_p^-  } 
\Big]
\nnb
\=  I \Big(  \frac{q_\para^2}{4m^2} \Big)
\label{eq:Pi_infinite-mu}
\, .
\end{eqnarray}
The above result agrees with the $ I $ function (\ref{eq:I}), 
and cancels the same function in the vacuum contribution (\ref{eq:vac_massive}). 
Therefore, the sum of the vacuum and thermal contributions reproduces 
the massless result (\ref{eq:Pi-massless}) in the infinite-density limit.

Finally, we make a comment on the infinite-temperature limit at zero density 
($T/m \to \infty  $ and $ \mu=0 $). 
In this limit, the Fermi distribution function becomes flat 
and just cuts off the momentum integral at the temperature scale. 
Thus, one can make a replacement $ p_F , \, \mu \to T $ in Eq.~(\ref{eq:Re-zero-T}) 
to arrive at the same expression (\ref{eq:Pi_infinite-mu}) as the infinite-density limit: 
\begin{eqnarray}
\lim_{T/m \to \infty} \left. \tilde \Pi_\para^{\temp }(\omega, q_z) \right|_{\mu=0}
\=  I \Big(  \frac{q_\para^2}{4m^2} \Big)
\label{eq:Pi_infinite-T}
\, .
\end{eqnarray}
Therefore, the sum of the vacuum and thermal contributions reproduces 
the massless result (\ref{eq:Pi-massless}) in the infinite-temperature limit.

\cout{
Finally, we make a comment on the infinite-temperature limit at zero density 
and confirm the disappearance of the pair-creation threshold. 
In this limit, the Fermi distribution function cuts off the momentum integral at the temperature scale. 
Thus, one can make a replacement $ p_F \to T $ in Eq.~(\ref{eq:Re-zero-T}). 
According to Eq.~(\ref{eq:Pi_zero-T}), we have 
\begin{eqnarray}
\lim_{T/m \to \infty} \left. \tilde \Pi_\para^{\temp }(\omega, q_z) \right|_{\mu=0}
\=  \frac{ m^2 \,  m_B^2 }{ q_\para^2\sqrt{1- 4m^2/q_\para^2 }  } 
\Big[ 
s_+ \ln  \frac{ p_z^+ -  \epsilon_p^+ }  { p_z^+ +  \epsilon_p^+  } 
-
s_- \ln  \frac{ p_z^- -  \epsilon_p^- }  { p_z^- +  \epsilon_p^-  } 
\Big]
\nnb
\= m_B^2 I \Big(  \frac{q_\para^2}{4m^2} \Big)
\label{eq:Pi_zero-mu}
\, ,
\end{eqnarray}
where we made use of the replacements $ p_F , \,  \mu \to T   $ 
and a relation, $ \arctan z = \frac{i}{2} \ln  \frac{ 1 - i z }{ 1 + i z} $. 
The result agrees with the $ I $ function (\ref{eq:I}), 
and cancels the same function in the vacuum contribution (\ref{eq:vac_massive}). 
Therefore, the sum of the vacuum and thermal contributions reproduces 
the massless result (\ref{eq:Pi-massless}) in the infinite-temperature limit, i.e., $T/m \to \infty  $. 

}

%

\bibliography{bib}

\end{document}

%% file: preamble.tex
\usepackage{graphicx}
\usepackage{latexsym}
\usepackage{amsfonts}
\usepackage{amssymb}
\usepackage{amsmath}
\usepackage{bm}
\usepackage{mathrsfs}
\usepackage{slashed}
\usepackage{ascmac}
\usepackage{comment}

\usepackage{dcolumn}

\allowdisplaybreaks[3]

\usepackage[
bookmarks=true,colorlinks,linkcolor=blue,urlcolor=cyan,citecolor=red]{hyperref}

\newcommand{\cout}[1]{ \if 0 {#1} \fi }

\newcommand{\beq}{\begin{eqnarray}}
\newcommand{\eeq}{\end{eqnarray}}
\newcommand{\bseq}{\begin{subequations}}
\newcommand{\eseq}{\end{subequations}}
\newcommand{\nn}{\nonumber}
\renewcommand{\=}{&=&}
\newcommand{\nnb}{\nonumber \\}

\newcommand{\sla}{ \slashed }

\renewcommand{\a}{\alpha}
\renewcommand{\b}{\beta}

\newcommand{\ep}{ \epsilon }

\newcommand{\gam}{ \gamma }

\newcommand{\bp}{{\bm{p}}}
\newcommand{\bq}{{\bm{q}}}

\newcommand{\bB}{{\bm B}}

\newcommand{\M}{ {\mathcal M} }

\newcommand{\order}{{\mathcal O}}
\newcommand{\prj}{ {\mathcal P} }

\newcommand{\para}{ \parallel}

\newcommand{\EM}{{\rm em}}

\newcommand{\tr}{ {\rm tr} }
\newcommand{\diag}{ {\rm diag} }
\newcommand{\sgn}{ {\rm sgn} }

\newcommand{\vac}{ {\rm vac} }

\newcommand{\LLL}{ {\rm LLL} }

\usepackage{color} 
\usepackage[normalem]{ulem}  

\newcommand{\cgd}[1]{{\color[rgb]{1,0,0}{#1}}}
